\documentstyle[11pt, aasms4]{article}
\input epsf
\begin{document}
\title{\bf Clustering of the Diffuse Infrared Light from the COBE DIRBE
maps. III. Power spectrum analysis and excess isotropic component of 
fluctuations.}

\author{A. Kashlinsky$^{1, 3}$, J. C. Mather$^2$, S. Odenwald$^4$
\\
$^1$NORDITA, Blegdamsvej 17, DK-2100 Copenhagen, Denmark\\
$^2$Code 685, NASA Goddard Space Flight Center, Greenbelt, MD 20771\\
$^3$Raytheon  STX, Code 685, \\ NASA Goddard Space
Flight Center, Greenbelt, MD 20771\\
$^4$Raytheon  STX, Code 630, \\ NASA Goddard Space
Flight Center, Greenbelt, MD 20771\\
}


\def\plotone#1{\centering \leavevmode
\epsfxsize=\columnwidth \epsfbox{#1}}

\def\wisk#1{\ifmmode{#1}\else{$#1$}\fi}

\def\wm2sr {Wm$^{-2}$sr$^{-1}$ }		
\def\nw2m4sr2 {nW$^2$m$^{-4}$sr$^{-2}$\ }		
\def\nwm2sr {nWm$^{-2}$sr$^{-1}$\ }		
\def\nw2m4sr {nW$^2$m$^{-4}$sr$^{-1}$\ }
\def\Ncut {$N_{\rm cut}$\ }
\def\lt     {\wisk{<}}
\def\gt     {\wisk{>}}
\def\le     {\wisk{_<\atop^=}}
\def\ge     {\wisk{_>\atop^=}}
\def\lsim   {\wisk{_<\atop^{\sim}}}
\def\gsim   {\wisk{_>\atop^{\sim}}}
\def\kms    {\wisk{{\rm ~km~s^{-1}}}}
\def\Lsun   {\wisk{{\rm L_\odot}}}
\def\Msun   {\wisk{{\rm M_\odot}}}
\def\um     {$\mu$m}
\def\sig    {\wisk{\sigma}}
\def\etal   {{\sl et~al.\ }}
\def\eg	    {{\it e.g.\ }}
\def\ie     {{\it i.e.\ }}
\def\bsl    {\wisk{\backslash}}
\def\by     {\wisk{\times}}
\def\cosec {\wisk{\rm cosec}}
\def\mic { $\mu$m \ }

\def\amin   {\wisk{^\prime\ }}
\def\asec   {\wisk{^{\prime\prime}\ }}
\def\cc     {\wisk{{\rm cm^{-3}\ }}}
\def\deg     {\wisk{^\circ}}
\def\ddeg   {\wisk{{\rlap.}^\circ}}
\def\damin  {\wisk{{\rlap.}^\prime}}
\def\dasec  {\wisk{{\rlap.}^{\prime\prime}}}
\def\approxeq{$\sim \over =$}
\def\abouteq{$\sim \over -$}
\def\percm{cm$^{-1}$}
\def\percmsq{cm$^{-2}$}
\def\percmcub{cm$^{-3}$}
\def\perhz{Hz$^{-1}$}
\def\perpc{$\rm pc^{-1}$}
\def\persec{s$^{-1}$}
\def\peryr{yr$^{-1}$}
\def\te{$\rm T_e$}
\def\tenup#1{10$^{#1}$}
\def\to{\wisk{\rightarrow}}
\def\thin{\thinspace}
\def\uk{$\rm \mu K$}
\def\p{\vskip 13pt}


\begin{abstract} The cosmic infrared background  (CIB) radiation is the
cosmic repository for energy release throughout the history of the
universe.  The spatial fluctuations of the  CIB resulting from galaxy
clustering are expected to be at least a few  percent on scales of a
degree, depending on the luminosity and clustering history of the  early
universe. Using the all-sky data from the COBE DIRBE instrument at
wavelengths 1.25 - 100 \um\ we attempt to measure the CIB fluctuations.
In the near-IR, foreground emission is dominated by small scale
structure due to stars in the Galaxy.   There we find a strong
correlation between the amplitude of the fluctuations  and Galactic
latitude after removing bright foreground stars. Using data outside the
Galactic plane ($|b| > 20^\deg$) and away from the center ($90\deg< l <
270\deg$) we extrapolate  the amplitude of the fluctuations to
cosec$|b|=0$.  We find a positive intercept of $\delta F_{\rm rms} 
=15.5^{+3.7}_{-7.0},
5.9^{+1.6}_{-3.7}, 2.4^{+0.5}_{-0.9},
2.0^{+0.25}_{-0.5}$
 \nwm2sr at 1.25, 2.2, 3.5 and 4.9 $\mu$m
respectively, where the errors are the range of 92\% confidence limits. For 
color 
subtracted maps between band 1 and 2 we find the 
isotropic part of the fluctuations at $7.6^{+1.2}_{-2.4}$ \nwm2sr .
Based on detailed numerical and analytic models,
this residual is not likely to originate from the Galaxy, our clipping
algorithm, or instrumental noise. We demonstrate that the residuals from the
fit used in the extrapolation are distributed isotropically and suggest
that this extra variance may result from structure in the CIB. 
We also obtain a positive intercept
from a linear combination of maps at 1.25 and 2.2 $\mu$m.   For $2^\deg
< \theta < 15^\deg$, a power-spectrum analysis  yields firm upper limits of
$(\theta/5^\deg) \times\delta F_{\rm rms} (\theta) < $ 6, 2.5, 0.8, 0.5
\nwm2sr at 1.25, 2.2, 3.5 and 4.9 $\mu$m respectively.  From 10 - 100
\um,   the dominant foregrounds are emission by dust in the Solar system
and the Galaxy. There the upper limits on the CIB fluctuations are below
1 \nwm2sr and are lowest ($\leq$ 0.5 \nwm2sr ) at 25 \um.
\end{abstract}


\section{Introduction}
\label{s1}
Diffuse backgrounds and their structure contain important information
about the history of the early Universe, when discrete objects either do
not exist or are not accessible to telescopic studies. The formation and
early evolution of galaxies should have generated radiation redshifted
into the infrared bands  (e.g. Partridge and Peebles 1967;  Bond et al.\
1986). This cosmic infrared background (CIB) may come from the entire
history of the Universe between the epoch of last scattering, mapped by
the microwave background, and the present day.

The COBE Diffuse InfraRed Background Experiment (DIRBE) (Boggess et al.\
1992; Silverberg et al.\ 1993;  Weiland et al., DIRBE Explanatory
Supplement, 1998) mapped the entire sky with 0.3\deg\ pixels and
0.7\deg\ resolution in ten bands  between 1.25 and 240 \um. The results
have been published (Hauser et al.\ 1998;  Kelsall et al. 1998;  Arendt
et al.\ 1998;  Dwek et al., 1998), based on a zodiacal light model
believed accurate within a few percent (Kelsall et al.\ 1998).  Their
Galactic dust emission model (Arendt et al.\ 1998) is derived from fits
to the DIRBE data and comparison to hydrogen maps, and accounts for the
variation of dust temperature.  Their  model for Galactic starlight is
derived from external star count models, with no free parameters. The
residuals from the modeling were not significantly above the
uncertainties except at 140 and 240 \um.   The FIRAS instrument on COBE
gave  similar answers, (Fixsen, Mather, Bennett, and Shafer, 1998),
using three different methods for removing the foregrounds. The
agreement  is significant because the instruments were calibrated
independently. The measured values of the far-IR background are
comparable to the total observed Galactic emission at visible and
near-IR wavelengths, and imply that about half of the luminosity of the
universe has been obscured by dust and converted into far-IR radiation.
Some distant galaxies might be more reddened or obscured by dust than
otherwise expected, and the CIB and its fluctuations might be brighter
than predicted from visible band galaxy counts.

In this paper we continue our search for a CIB by trying to detect its
spatial structure in the DIRBE data. The method is similar to that first
suggested by Gunn and later applied in the visible (Shectman 1973, 1974)
and UV (Martin and Bowyer 1989). Recently, Vogeley (1998) applied the
method to the Hubble Deep Field constraining the amount of the visible
cosmic background,  concluding that there is very little visible cosmic
background  from unresolved sources. They considered arcsecond scales
where fluctuations are dominated by random parts of individual galaxy
(Vogeley 1998) or cluster (Shectman 1974) profiles. We apply the method
to degree scales where the dominant contribution comes from galaxy
clustering. Our previous results  are described by Kashlinsky, Mather,
Odenwald and Hauser (1996a;  Paper I) and Kashlinsky, Mather and
Odenwald (1996b;  Paper II).

If we assume that the galaxies producing the CIB are  clustered, the
amplitude of the resulting CIB 2-point correlation function,
$C(\theta)$, depends on the rate of the flux emission, $dF/dz$, and on
the galaxy 2-point correlation function, $\xi(r; z)$. If the latter is
known on the linear scales subtended by a given angle, measurements or
upper limits on  $C(\theta)$ can constrain the levels of the CIB emitted
by clustered material. Measuring or limiting the structure of the CIB
can be especially valuable in the mid- to far-IR bands  where the
foreground emission is very bright,  but smooth.

The plan of the paper is as follows: In Sec.\ \ref{s2} we define the
quantities used  and provide mathematical background.  We show that for
the measured galaxy correlation function we expect  fluctuations of
about 5 to 10\% in the CIB flux on 0.5\deg\ scales, almost independently
of the particular mechanism of the CIB production. We discuss our search
for the CIB structure in the DIRBE  data between 1 and 100 \um. The last
two bands  of DIRBE - 140 and 240 \um\  - are not useful for this
because of the large instrument noise there. Section \ref{s3} deals with
the beam profile and the spatial window function for power spectrum
analysis. Section \ref{s4} discusses the data for $\lambda \leq 5$ \um,
where the foregrounds are dominated by stellar emission from the Galaxy.
We find a significant trend in the amplitude of the measured
fluctuations with Galactic latitude. When our star removal algorithm is
applied to simulated data, we recover the measured slope of the
correlation with Galactic latitude, and  extrapolation of the simulated
fluctuations to $\cosec|b|=0$  leads to a zero intercept. We derive a
formula relating the measured fluctuations to star counts  and show that
for a plane-parallel Galaxy there should be a simple relationship
between the $b$ dependence of the fluctuations and the slope of the star
counts function at the Galactic pole.  Then we show that for $ 90\deg <
l < 270\deg$ and $|b|>20\deg$ the measured structure of the Galaxy fits
the  plane-parallel model.  Furthermore, the scatter in the $C(0)$ -
cosec$|b|$  plot is low enough to allow an extrapolation to
$\cosec|b|=0$, with positive  intercepts in all four near-IR bands. We
perform a power spectrum analysis but most of the measured structure is
due to fluctuating star counts. If the star removal algorithm removes
stars fainter than 5 times the confusion noise limit, there are too few
pixels left to calculate a power spectrum. Nevertheless, limits are
sufficiently low to be of interest in testing theoretical predictions.
We test our methods with numerical and analytic models of the Galaxy and
possible CIB contributions and find  good agreement with the data.
Results from the mid- to far-IR bands, where foreground emission
structure is dominated by the large scale gradients from dust emission
in the Solar System and the Galaxy, are presented in Sec.\ \ref{s5}. The
upper limits on these fluctuations are $<$ 1 \nwm2sr , depending on
wavelength, and provide much stronger tests of the predictions of galaxy
counts than do direct measurements from inside the zodiacal dust cloud.
We summarize our results in Sec.\ \ref{s6}.

\section{Theoretical preliminaries} \label{s2}

Here we provide a general mathematical basis without specific
cosmological and galaxy evolution models. We will use the data to
constrain the CIB properties in a model-independent way, and then to
constrain  models of  galaxy evolution. We  extend our previous work (Paper 
I,I; 
Jimenez \& Kashlinsky 1999) to
show the power spectrum of the fluctuations and to estimate the typical
amplitude of the fluctuations in a model-independent way.

We start with definitions. The surface brightness in the CIB per unit
wavelength will be denoted as $I_\lambda$, per unit frequency as
$I_\nu$, and per logarithmic wavelength interval $F=\lambda I_\lambda =
\nu I_\nu$, and we call them all ``flux.''  The fluctuation in the CIB
flux is then $\delta F(\mbox{\boldmath$x$})= F(\mbox{\boldmath$x$}) -
\langle F \rangle$, where $\mbox{\boldmath$x$}$ is the two dimensional
coordinate on the sky and $\langle ... \rangle$ denotes ensemble
averaging. We also use the two-dimensional Fourier transform, $\delta
F(\mbox{\boldmath$\theta$})= (2\pi)^{-2} \int \delta F_q
\exp(-i\mbox{\boldmath$q$}\cdot \mbox{\boldmath$\theta$})
d^2\mbox{\boldmath$q$}$.

If  $\delta F(\mbox{\boldmath$x$})$ field is a random variable, then it
can be described by the moments of its probability distribution
function. The first non-trivial moment is the  {projected 2-dimensional}
correlation function $C(\theta) = \langle \delta
F(\mbox{\boldmath$x$}+\theta) \delta F(\mbox{\boldmath$x$})\rangle$. The
2-dimensional power spectrum is $P_2(q) \equiv \langle |\delta
F_q|^2\rangle$, where the average is performed over all phases. The
correlation function and the power spectrum are a pair of 2-dimensional
Fourier transforms and for isotropically distributed signal are related
by
\begin{equation}
C(\theta)= \frac{1}{2\pi} \int_0^\infty P_2(q) J_0(q\theta) q dq,
\label{e1}
\end{equation}
\begin{equation}
P_2(q)= 2\pi \int_0^\infty C(\theta) J_0(q\theta) \theta d\theta,
\label{e2}
\end{equation}
where $J_n(x)$ is the $n$-th order cylindrical Bessel function. If the
phases are random, then the distribution of the flux field is Gaussian
and the correlation function, or its Fourier transform the power
spectrum, uniquely describe its statistics. In  measurements with a
finite beam radius $\vartheta$ the intrinsic power spectrum is
multiplied by the window function $W$ of the instrument. Another useful
quantity is the mean square fluctuation within a finite beam, or
zero-lag correlation signal, which is related to the power spectrum by
\begin{eqnarray}
\langle (\delta F)^2 \rangle_\vartheta = \frac{1}{2\pi} \int_0^\infty P_2(q)
W_{TH}(q\vartheta) q dq \nonumber \\
\sim \frac{1}{2\pi} q^2P_2(q)|_{q\sim {\pi}/{2}
\vartheta} .
\label{e3}
\end{eqnarray}
For  a top-hat beam the window function is $W_{TH}=[2J_1(x)/x]^2=0.5$ at
$x\simeq \pi/2$  where $x=q \vartheta$,  and hence the values of
$q^{-1}$ correspond to fluctuations on angular scales of diameter $\sim
\pi/q$.

The CIB flux and its structure are measured in projection on the
celestial sphere and reflect both the 3-dimensional clustering pattern
of the galaxy distribution and the rate of emission at redshift $z$. We
introduce the 3-dimensional two-point correlation function of galaxy
clustering, $\xi(r)$, and its 3-dimensional power spectrum, $P_3(k)$.
These are related via 3-dimensional Fourier transforms, and assuming
isotropy are related by
\begin{equation}
\xi(r) = \frac{1}{2\pi^2} \int_0^\infty P_3(k) j_0(kr) k^2 dk .
\label{e4}
\end{equation}

The projected CIB correlation function is related to the underlying
two-point correlation function of the galaxy distribution and the rate
of the CIB flux emission via the Limber equation
\begin{equation}
C(\theta) = \int_{z_1} \int_{z_2} \frac{dF}{dz_1}\frac{dF}{dz_2}
\xi(r_{12}; z) dz_1dz_2,
\label{elimber}
\end{equation}
where $r_{12}$ is the {proper} length subtended by the angle $\theta$
and redshifts $z_1, z_2$. In the limit of small angles, $\theta \ll 1$,
the Limber equation becomes (e.g. Peebles 1980)
\begin{equation}
C(\theta) = \int_0^\infty dz \left(\frac{dF}{dz}\right)^2 \int_{-\infty}^\infty
\xi(r_{12}; z) du.
\label{e5}
\end{equation}
For  Robertson-Walker metrics the proper separation is given by
\begin{eqnarray}
r_{12}^2 =
(\frac{dx/dz}{
\sqrt{1-kx^2} } u)^2+\frac{x^2(z)\theta^2}{(1+z)^2}\\
 = c^2(\frac{dt}{dz})^2
u^2 +d_A^2(z)\theta^2.
\end{eqnarray}
Here $d_A(z)$ is the
angular diameter distance, $x(z)\equiv d_A(z)/(1+z)$, $u$ is the  integration
variable and $t$ is the cosmic time.

The comoving volume occupied by a unit solid angle in the redshift
interval $dz$ is $dV/dz = (1+z) x^2(z) cdt/dz$, and the power received
per unit wavelength and collecting area in band $\lambda$ from each
galaxy with absolute bolometric luminosity $L$ at redshift $z$ is
$[L/({4\pi x^2(1+z)^3})]f_\lambda(\frac{\lambda}{1+z}; z)$. Here
$f_\lambda d\lambda$ is the fraction of the total light emitted in the
wavelength interval $[\lambda;  \lambda +d\lambda]$ and the extra factor
of $(1+z)$ in the denominator accounts for the fact that the flux
received in band $\lambda$ comes from a redshifted galaxy. The
contribution to the total CIB flux from the redshift interval $dz$ is
given by
\begin{equation}
\frac{dF}{dz} = \frac{R_H}{4\pi}
\frac{1}{(1+z)^2} \frac{d(H_0t)}{dz}
\sum_i {\cal L}_i(z) [\lambda f_{\lambda, i}(\frac{\lambda}{1+z}; z)],
\label{e6}
\end{equation}
where the sum is taken over all galaxy populations contributing  flux in
the observer rest-frame band at $\lambda$, and $f_\lambda$ characterizes
the spectral energy distribution (SED) of galaxy population $i$. Here
$R_H=cH_0^{-1}$ and ${\cal L}(z)= \int \Phi(L; z) L dL$ is the comoving
luminosity density from galaxies with the luminosity function $\Phi(L;
z)$ at epoch $z$.

It is illustrative to study the redshift dependence of the flux rate
production, Eq.\ (\ref{e6}). At small redshifts the factor
$(1+z)^{-2}dt/dz$  varies little with $z$, and the rate of flux
production is governed by the comoving bolometric luminosity density
${\cal L}(z)$ and the SED of the galaxy emission $f_\lambda$. If the
luminosity evolution at these redshifts is small, the rate of flux
production is governed by the SED shape. If $f_\lambda(\lambda)$
increases towards shorter wavelengths then $dF/dz$ increases with $z$.
For $\lambda f_\lambda=$ const, and no luminosity evolution, the rate is
roughly constant with small $z$. In addition, there is observational
evidence for an increase of ${\cal L}(z)$ out to $z\simeq 1$ in the galaxy 
rest-frame UV
to near-IR (1 \um ) bands (Lilly et al.\ 1996). At sufficiently high redshifts,
the evolution in Eq.\ (\ref{e6}) would be offset by the factor
$(1+z)^{-2}dt/dz$, so that the rate of production would be cut off at
sufficiently large $z$. This factor is responsible for resolving Olbers'
paradox even for a flat SED.

The SED for rest-frame $\lambda < 10$\um\ is dominated by stellar
emission, with a peak at visible wavelengths and a decrease for $\lambda
> 0.7$ \um. Consequently, most of the predicted J band CIB comes from
redshifts $z\sim 0.3-1$, which shifts the visible emission of normal
stellar populations into the J band;  cf.\ Yoshii and Takahara (1988).
In the M band at 5 \um, most of the predicted CIB comes from $z>1-2$. At
$\lambda > 10$  \um, the emission is dominated by galactic dust and the
situation is reversed, so $f_\lambda$ increases with wavelength roughly
as $\lambda^\alpha$ with $\alpha \sim 1.5$. Hence, the dusty star-burst
galaxies observed by IRAS at low redshifts should make the dominant
contribution to the 10 \um\ CIB. In the far-IR, the measured CIB found
by DIRBE can have large contributions from high redshifts.

Measurements of the correlation function $C(\theta)= \langle \delta
F({\mbox{\boldmath$x$}} +\theta) \delta F({\mbox{\boldmath$x$}})
\rangle$ have an important
advantage over direct determinations of the CIB spatial power spectrum,
because they are immune to the discontinuities in the maps created by
point source removal. (Because, in practice, $C(\theta)$ is evaluated from 
masked data, the uncertainties of the $C(\theta)$ points are themselves 
correlated.) However, interpreting $C(\theta)$ in terms of
$dF/dz$ and $\xi(r)$ can be cumbersome  because the right-hand-side of
(\ref{e5}) contains a double integral and the 3-dimensional correlation
function is not always positive. By definition,  $\xi(r)$  must be
negative on large scales so that $\int_0^\infty \xi(r) r^2 dr=0$.

A simpler method is to work with the Fourier transform of the
correlation function, $P_2(q)$, which contains the same information as
$C(\theta)$ and is easier to interpret. As for $C(\theta)$ the effects of the 
mask will produce correlations of the power spectrum points with each other. 
Also, the measured power spectrum is the convolution of the Fourier transform 
of the mask with the true power. Although the uncertainties in $C(\theta)$ and 
$P_2(q)$ can be evaluated from theory, we evaluated them from the data by comparing 
multiple measurements.

The Limber equation (\ref{e5})
can be rewritten directly in terms of the power spectra, substituting
Eq.\ (\ref{e4}) into (\ref{e5}) and using $\int_{-\infty}^\infty
j_0(\sqrt{x^2 +y^2})dy = \pi J_0(x)$ to obtain
\begin{eqnarray}
C(\theta) = \frac{1}{2\pi} \int_0^\infty dz \left( \frac{dF}{dz} \right)^2
\frac{1}{cdt/dz} \nonumber \\
\int_0^\infty P_3(k; z) J_0(k d_A \theta) k\ dk .
\label{e7}
\end{eqnarray}
Substituting this expression into (\ref{e2}) and using the orthogonality
relations for the Bessel functions,\\ $\int_0^\infty J_0(\alpha \theta)
J_0(\beta \theta) \theta d\theta = \alpha^{-1} \delta_D(\alpha-\beta)$,
leads to
\begin{equation}
P_2(q) = \int_0^\infty
\left(\frac{dF}{dz}\right)^2 \frac{(1+z)^2}{c\frac{dt}{dz} x^2(z)}
P_3(qd_A^{-1}(z); z) dz.
\label{e8}
\end{equation}
Equation(\ref{e8}) involves only one integration and the kernel is
always positive. Hence it may be easier to derive cosmologically
interesting quantities such as $F$ and $\xi(r)$ or $P_3(k)$. Determining
$P_2(q)$ directly from the data is difficult owing to the masking
created by point source removal, and it seems best to compute
$C(\theta)$ and then derive $P_2(q)$ from it.

Figure  \ref{f1}a plots the argument, $q(1+z)/x(z)$, of $P_3$ in the
right-hand-side of Eq.\ (\ref{e8}) for the largest wavenumbers (smallest
scales) probed by DIRBE at various redshifts $z$. The dotted line
corresponds to $\Omega=1, \Omega_\Lambda=0$;  the solid to $\Omega=0.1,
\Omega_\Lambda=0$ and the dashes  to $\Omega=0.1, \Omega_\Lambda=0.9$,
where $\Omega_\Lambda=3H_0^2(1-\Omega)$ denotes the contribution of the
cosmological constant. For  redshifts contributing most of the CIB flux
($z>$0.1-0.2, cf. Paper I), the DIRBE instrument  probes scales which
are in the quasi-linear or linear regime, and which thus can be
approximated as having evolved at the same rate with time. Thus for
DIRBE scales, one can write $P_3(k; z) \simeq P_3(k; 0) \Psi^2(z)$,
(e.g. Peebles 1980), where $\Psi^2(z)$ accounts for the evolution of the 
clustering 
and is
normalized to $\Psi^2(0)=1$. Then Eq.\ (\ref{e8}) can be rewritten in a
more compact way:
\begin{equation}
q^2P_2(q)= \int_0^\infty
\left(\frac{dF}{dz}\right)^2 \frac{\Psi^2(z)}{H_0\frac{dt}{dz}}
\Delta^2(\frac{q}{d_A(z)}) dz,
\label{e9}
\end{equation}
with
\begin{equation}
H_0\frac{dt}{dz} =\frac{1}{(1+z)^2\sqrt{1+\Omega z +
\Omega_\Lambda[(1+z)^{-2}-1]}} .
\label{e10}
\end{equation}
The left-hand-side of (\ref{e9}) is the same order of magnitude as the
mean square fluctuation (\ref{e3}), and  we have defined
\begin{equation}
\Delta^2(k) \equiv R_H^{-1}k^2 P_3(k; 0).
\label{e11}
\end{equation}
The quantity $\Delta(k)$ is roughly the fluctuation over the
line-of-sight cylinder of length $R_H$ and diameter $k^{-1}$. For
relevant scales and spectra of density fluctuations, $\Delta(k)$
increases with $k$. Hence, the CIB fluctuation on a scale of $k^{-1}$ is
$\sim \sqrt{k^2 P_2(k)} \sim F \Delta(k/R_H)$.

The present day power spectrum of galaxy clustering has been measured on
scales corresponding to at least the smallest angular scales probed by
DIRBE. The most accurate measurement comes from the APM survey data on
the projected angular correlation function (Maddox et al.\   1990).
Baugh and Efstathiou (1993) deprojected the APM data to obtain the
underlying power spectrum of galaxy clustering, $P_3(k)$. Kashlinsky
(1998) used the current large scale data, and the abundance of objects
at high redshifts, to reconstruct the pregalactic density field over six
orders of magnitude in mass.  His result  requires significant
fluctuation power on small scales, an early epoch of galaxy formation,
and  high levels of the CIB and its fluctuations. Fig.\  \ref{f1}b plots
the data from the Baugh and Efstathiou (1993) deprojection. On small
scales (large $k$), $\Delta^2(k) \propto k^{0.7}$, so  the integrand in
Eq.\ (\ref{e9}) behaves as $\propto z^{-0.7}$ with an integrable
singularity at $z\rightarrow 0$. While a non-negligible part of the
clustering part of the CIB fluctuations  comes from nearby galaxies,
much of it arises from galaxies at  $z>1$.

In Papers I and II we estimated $\sim$ 5-10\% CIB fluctuations on a
scale of 0.5\deg. One can also see this in a more intuitive way from
Eq.\ (\ref{e9}). Fig.\ \ref{f1}a plots $q(1+z)/x(z)$, the largest
wavenumber that enters on the right hand side of (\ref{e9}), for
$q^{-1} = 0.5$\deg. In the Friedman-Robertson-Walker Universe,
$q/d_A(z)$ reaches a minimum at $z\sim 1-3$, and its value at the
minimum depends weakly on cosmological parameters. Thus (\ref{e9}) can
be rewritten as an inequality:
\begin{equation}
q^2P_2(q) \geq \Delta^2\left(\min[\frac{q}{d_A(z)}]\right) \int_0^\infty
\left(\frac{dF}{dz}\right)^2 \frac{\Psi^2(z)}{H_0\frac{dt}{dz}} dz .
\label{e12}
\end{equation}
On linear and quasi-linear scales over the range of redshifts that
contribute most to the CIB, the quantity $\Psi^2(z)/\frac{H_0dt}{dz}$
depends weakly on $z$ (e.g. on linear scales $\Psi^2=(1+z)^{-2}$ if
$\Omega=1$). The integral is of the same order of magnitude as $F^2$
because the term $dF/dz$ is a peaked function with a full width at half
maximum of order unity.  Hence,  the relative  fluctuation in the CIB is
$\sim \Delta(\min[\frac{q}{d_A(z)}]) $. Fig.\ \ref{f1}b plots
$\Delta(k)$ vs.\ $k$  using the power spectrum of APM galaxies from
Baugh \& Efstathiou (1993); it is an increasing function of  $k$.  At
the  wavenumber of the minimum in Fig.\ \ref{f1}a, the CIB fluctuation
can  be as large as $\sim$ (5-10)\%. In models where the bulk of the CIB
comes from higher redshifts, this number may be somewhat smaller.
Because this method measures a two-point process, the results constrain
a measure related to the mean square of the CIB emission rate, i.e.\
$\sim \int \left(\frac{dF}{dz}\right)^2 dz$.

Many models have calculated the expected CIB over the 1 - 100 \um\ range  (e.g.
Partridge and Peebles 1967;  Stecker et al.\ 1977; Bond et al.\   1986;
Fall et al.\   1996; Wang 1991; Beichman and Helou 1991; Franceschini et
al.\   1991;  Cole et al.\   1992; Malkan and Stecker, 1997;  Jimenez
and Kashlinsky 1999;  Dwek et al.\   1998). The models are normalized to
galaxy counts and predict a typical  flux of  $F \sim 10$ \nwm2sr . Deep
K band counts of galaxies  (e.g. Cowie et al.\   1994; Djorgovski et
al.\   1995) or at 12 - 100 \um\ (Hacking and Soifer 1991) suggest
{minimal} fluxes of at least a few \nwm2sr . Therefore the CIB
fluctuations on 0.5\deg\ scales may be $\sim 1$ \nwm2sr .

The APM measurements  of the galaxy correlation function in the blue
band might not apply to the infrared. However, the $r$ band  Palomar
survey (Picard 1991), and the R-band  Las Campanas survey (Shectman
et al. 1996, Lin et al. 1996) give results identical to the APM survey. On the 
other
hand,  for IRAS galaxies  on small (non-linear) scales the  correlation
function has a lower coherence length.  Saunders et al.\ (1992) show
that on small scales IRAS galaxies at 60\um\  cluster with $\xi_{IRAS} =
(r/r_{*, IRAS})^{-1-\gamma_{IRAS}}$ where $r_{*, IRAS} \simeq
4h^{-1}$Mpc and $\gamma_{IRAS}\simeq 0.6$, as opposed to
$r_*=5.5h^{-1}$Mpc and $\gamma\simeq 0.7$ for the APM galaxies (Moore et
al.\   1994). The slight difference could be due a tendency of the IRAS
dusty star-burst galaxies to avoid the central regions of rich clusters
of galaxies. However, on larger (linear) scales where the galaxy
clustering pattern presumably traces the pregalactic density field, IRAS
galaxies as measured by the QDOT counts-in-cells analysis (Saunders et
al.\   1990) are consistent with the APM galaxies power spectrum. These
linear to quasi-linear scales are relevant for the DIRBE beam size, so
we use the APM numbers for all infrared wavelengths. In surveys with
smaller beams, the differences will be more pronounced, but the
differences between the IRAS and APM correlation functions do not lead
to appreciably smaller mid- to far-IR CIB fluctuations.

In addition to the clustering term, there is a white-noise component in
the correlation signal due to individual galaxies (e.g.\ Peebles 1980).
Its amplitude at zero lag is
\begin{eqnarray}
C_{\rm WN}(0) = \theta_{\rm beam}^{-2}
\frac{R_H}{16\pi^2} \int \frac{\int \Phi(L; z) L^2 dL}{x^2(z)(1+z)^8}
\frac{H_0dt}{dz} \nonumber \\
\left[ \lambda f_\lambda (\frac{\lambda}{1+z}; z) \right]^2 dz .
\label{e13}
\end{eqnarray}
Like white noise from discrete stars, the white noise from galaxies is
dominated by the nearest objects, unlike the clustered component
(\ref{e5}), because of the presence of $x^2(z)$ in the denominator of
the integrand in (\ref{e13}). Because the galaxies are almost
undetectable against the star fluctuations, the galaxy white noise is
also negligible.

\section{DIRBE data and beam profile} \label{s3}

The  COBE DIRBE instrument  provided an all-sky 41 week survey with a
ten-band photometer (Boggess et al.\ 1992). The DIRBE bolometer
measurements at  140 and 240 \um\ are too noisy for our purposes. The
remaining eight bands are centered on wavelengths 1.25, 2.2, 3.5, 4.9,
12, 25, 60 and 100 \um, and are labeled Bands 1-8 respectively. We
obtained the 41 weekly-averaged DIRBE maps from the National Space
Science Data Center (NSSDC), subtracted the zodiacal light model
developed by the DIRBE team (Kelsall et al. 1998; Weiland et al, 1998,
DIRBE Explanatory Supplement), and  averaged the weeks together.

The maps are stored in a cube format and pixelized into 6 faces of
256$^2$ approximately square pixels $\sim 0.3$\deg\  on a side. For a
finite beam the ideal map is convolved with the beam window function.
For  a circular top-hat beam similar to DIRBE's square top hat, the
Fourier transform of the window function is $W(x) = [2J_1(x)/x]^2$,
where $J_1(x)$ is the first-order cylindrical Bessel function. The
measured power spectrum is then the product of the underlying power
spectrum and the beam window function, i.e. $P_{\rm measured}(q) =
P_2(q) W(q\vartheta)$ where $\vartheta$ reflects the beam size. On large
scales (small $q$), $W(x)\sim 1$. Throughout the rest of the paper,
$P(q)$ will refer to the 2-dimensional power spectrum of the diffuse
emission computed from DIRBE maps after deconvolution from the beam
profile, i.e. $P(q)= P_{\rm measured}(q)/W(q\vartheta)$.

We determined the effective $W$ and the effective beam size from  maps
of the beam response function archived at the NSSDC with 181$\times$225
0.0065\deg\ square pixels, which we embedded in 256$^2$, 512$^2$, and
1024$^2$ pixel arrays. The results for Band 1 are shown in Fig.\
\ref{f2}, and are similar for all eight bands. The solid line shows a
top-hat beam profile with $\vartheta=0.4$\deg\  which is slightly lower
than the value of $\vartheta=0.45$\deg\  adopted in Paper I.   From  eye
fits we adopted $\vartheta=0.4$\deg\  in Band 1, 0.37\deg\  in Bands 2
and 3, and 0.35\deg\ in Bands   4 to 8. The combined effects of beam
smearing from the motion of the beam during sampling, pixelization, and
pointing error would increase the effective beam size by about 10\%, and
agree better with Paper I. The beam response function drops below 10\%
at $q^{-1} <$0.15\deg\ and the pixelization prevents measurement below
$q^{-1}$ = 0.1\deg.

\section{Near-IR analysis} \label{s4}

\subsection{Foregrounds and point source removal} \label{ss41}

Foreground emission from the Galaxy and the solar system is the main
problem in unveiling the expected CIB.  At wavelengths less than 10 \um,
the dominant foreground after removing the zodiacal light model is
emission from stars in our Galaxy. Using the SAO Bright Star Catalog
magnitudes and colors, we simulated J band maps.  We found that aside
from the large-scale shape of the galaxy and a few star clusters this
foreground has an almost uncorrelated spatial distribution. At 12 and 25
\um\  (DIRBE Bands   5 and 6) the zodiacal dust is so bright that
residuals from the model subtraction dominate the map structure outside
the Galactic plane. At wavelengths of 60 and 100 \um\  (DIRBE bands 7
and 8) most foreground residual fluctuations come from cirrus dust
clouds in the Galaxy, and stars contribute little.

The measured fluctuations contributed by point sources  can be reduced
substantially by identifying and removing them down to near the
confusion noise limit. We used the point source finding routine
developed by  the DIRBE team and adopted in Paper I, and call it the
``clipping'' algorithm.  The data from a selected region (patch) are
first used to construct a smoothed model of the sky background emission
for the whole patch, as follows.  Surrounding each pixel,  a window of
size $f_{\rm size}\times f_{\rm size}$ is searched for the minimal flux
value. We use the minimal value rather that the median because the stellar 
brightness distribution is highly skewed. The map of this minimal (or lower 
envelope) flux is then fitted
to a 2-dimensional polynomial surface of order $n_{\rm fit}$. This
polynomial is in turn subtracted from the sky map, and the standard
deviation $\sigma_0$ is calculated. Then, each of  the pixels with flux
above $N_{\rm cut}\sigma_0$ is masked out along with  the  8 adjacent
pixels, about twice the DIRBE beam area. This process of fitting a
polynomial, identifying bright pixels,  and masking them out, is
iterated about 5 times,  until no more fluxes above \Ncut$\sigma_0$ are
found.  In the near-IR no noticeable improvement is reached above
$n_{\rm  fit}=2$. We discuss the dependence on $f_{\rm size}$ in Sec.\
\ref{ss45}; for the results  presented below we used $f_{\rm size}=5$.
The polynomial fit was used only for point source masking,  no gradients
were subtracted from the maps analyzed in this (near-IR) section, and
the results at high galactic latitude ($|b|>20$\deg) were independent of
this choice.

For the near-IR, where foreground stars are the dominant source of
fluctuations, there should be a clear correlation between the fluxes at
different wavelengths. Furthermore, because most of the flux and
fluctuations come from K and M stars, and the DIRBE pixels are large
enough to include many stars, the dispersion in the color diagrams for
the four near-IR bands  should be small. In Paper I we  used these
properties to search for CIB fluctuations that have a different color
due to redshifts or a different source spectrum. We construct a linear
combination of maps at two different bands:
\begin{equation}
\Delta_{12} = \delta F_1 -\beta_{12} \delta F_2 .
\label{e14}
\end{equation}
Note that this $\Delta$ is unrelated to the $\Delta$ of Eq.\ (\ref{e9}).
The variance of this map, $C_\Delta(0)\equiv \langle
\Delta_{12}^2\rangle$, is minimized for
\begin{equation}
\beta_{12} = \frac{ \langle \delta F_1\delta F_2\rangle }{ \langle \delta
F_2^2\rangle}.
\label{e15}
\end{equation}
Because the dispersion in the color index $\beta = \delta F_1/\delta
F_2$ is  small, i.e. $\sigma_\beta = \sqrt{\langle \beta^2 \rangle
-\langle \beta \rangle^2 } \ll \beta_{12}^2$ with the average taken over
all the pixels in the patch, the foreground contribution to $C(0)$ will
be reduced by a factor $\sim (\beta_{12}/\sigma_\beta)^2 \gg 1$. If the
fluxes in adjacent bands  do not correlate,  the emission in the two
bands  comes from different sources or from measurement errors (noise).

Since most of the predicted CIB comes from galaxies at $z>0.2$ with
typical stellar populations, its color should differ from that of the
foreground stars, and the color subtracted maps should retain some of
the CIB structure. If most of the CIB comes from high redshifts so that
\begin{equation}
\beta_{CIB}=\langle \delta F_{1, CIB}
\delta F_{2, CIB}\rangle/\langle \delta F_{2, CIB}^2\rangle \leq 2
\beta_{12},
\end{equation}
the CIB fluctuations in the color subtracted maps will be larger than in
the single band maps. We used Eq.\ (\ref{e15})  to make all-sky
color-subtracted maps for all the adjacent band pairs, in order to search for a
coordinate-independent part of the fluctuations. This method can be
generalized to a multi-color subtraction method, e.g.\ minimize
$\Delta_{123} = \delta F_1 -\beta \delta F_2 - \alpha \delta F_3$ with
respect to $\alpha, \beta$. We applied this to the DIRBE maps, but
without significantly different results.

\subsection{All sky variance analysis: $C(0)$} \label{ss42}

We divided the sky into 384 patches, each with $32\times 32$ pixels, and
clipped each patch individually using the above procedure, with  \Ncut =
7, 5, 3.5 and 3. Because the foreground emission at these wavelengths is
dominated by point sources, very few pixels are left for \Ncut$<3$. We
performed the same analysis on 96 patches of $64\times 64$ pixels, with
similar results.

Our star finding algorithm cuts deeper into the distribution than a
simple interpretation might suggest.  Each bright star is observed in
about 5 pixels, depending on the position of the star. The noise
distribution of these pixels in the absence of the bright star still has
a non-Gaussian distribution with a non-zero mean.  Therefore a star near
the clipping threshold will be identified in the pixel where the
background noise fluctuations are greatest, rather than at the true star
location.  Hence, the clipping algorithm finds stars about 1 $\sigma$
fainter than \Ncut, as we confirmed with simulations.

After clipping the 384 patches, we computed the color indices according
to Eq.\ (\ref{e15}), the single band $C(0)$, and the color-subtracted
$C_\Delta(0)$ for each patch. Figure  \ref{f3} shows  histograms of the
numbers of the remaining pixels after clipping. In the near-IR bands,
about 350-450 pixels out of 1024, or about uncorrelated 75-90 beams per patch, 
remain
for $N_{\rm cut}=3.5, 3$, making the intrinsic uncertainty of $C(0)$ less
than 15\%.

In each patch there is a clear correlation between the fluxes in the
adjacent bands.  Paper I gave flux correlation plots for selected
patches and an earlier model of the zodiacal light. The current data
give similar results. For bands 1-4 all the patches have correlation
coefficient $R >0.9$ between all the pairs of  bands. The color indices
have very small dispersion, $\sigma_\beta/\beta < 10\%$.  Because the
dispersion in $\beta$ is so small, and most of the predicted CIB
emission at these wavelengths comes from redshifts $z > 0.1$, the CIB
fluctuations should not be removed by the color subtraction.

Fig.\ \ref{f4} plots the histograms of $\beta$ for  $N_{\rm cut}=3$.
The maps for the band pair [4, 5] have color index $\beta \simeq 0$
because the dust in the solar system is not strongly correlated with the
stars seen in Band 4. Between Bands  1 and 2, most of the patches have a
color index of $\beta_{12} \simeq 2$,  typical of K-M giants. In band
pairs [2, 3] and  especially [3, 4] the spread in $\beta$  is
substantial, suggesting that star fluctuations are not the only source,
or that their color is different in different regions. The range of
color indices for the band differences $[2-3]$ and $[3-4]$ is reduced
for $|b|>20$\deg, but  nevertheless remains much wider than for the
$[1-2]$ maps. Most of the map peaks are due to stars, as we found  them
in simulated maps derived from the SAO Bright Star Catalog, using
wavelength extrapolations appropriate to the tabulated  spectral types.

The lowest limits on $C(0)$ for single bands are similar to those in
Paper II, and for the color subtracted bands they are not very different
from Paper I. However, we can now address the dependence of $C(0)$ on
Galactic coordinates. Fig.\ \ref{f5} shows a strong correlation of
$\sqrt{C(0)}$ with $\cosec |b|$. Fig.\ \ref{f6} plots $\sqrt{C(0)}$ vs.\
cosec$|b|$ for the color subtracted maps [1-2], [2-3], and [3-4].
Although the color subtracted $C(0)$  is  a factor of $\sim 10$ lower
than for single bands, the residual fluctuations still depend strongly
on Galactic latitude.

\subsection{Near-IR fluctuations from Galactic stars} \label{s43}

We would now like to extrapolate the data to estimate the extragalactic
contribution to the fluctuations.  We use analytic and numerical models
to support a power-law dependence of the stellar contribution to $C(0)$
on cosec$|b|$.  We also show how to extrapolate the fluctuations due to
Galactic stars to cosec$|b|=0$, and that we expect no isotropic residual
for $C(0)$ due to stars.

We assume that Galactic stars are distributed with a white noise power
spectrum, i.e.\ they are spatially uncorrelated except for large scale
distribution functions such as those in the Bahcall and Soneira (1980)
and Wainscoat et al.\ (1992) models.  Thus $C(0)$ in the direction $b$
is given by the following (Galactic) version of the Limber equation:
\begin{eqnarray}
\sigma^2(x) = \sum_i \int \frac{L_1}{4\pi r_1^2}\frac{L_2}{4\pi r_2^2} dP_{12}
\nonumber \\
= \sum_i \frac{L_i^2}{4\pi} \omega \int_{R_i}^\infty \frac{n_i(r;x)}{r^2} dr.
\label{e16}
\end{eqnarray}
Here the sum is taken over  stars of type $i$ with intrinsic luminosity
$L_i$, $\omega$ is the pixel solid angle, $x = {\rm cosec}|b|$, $r$ is
the distance to each star, and $n_i$ is the number density of stars at
that distance in the direction of $b$. The distance to the star is
related to the height, $Z$, above the Galactic plane by $r=
Z$cosec$|b|$. In our clipping algorithm, we remove sources exceeding the
flux limit of \Ncut$\times$(standard deviation of the background
polynomial fit). This standard deviation is approximately $\sigma$ from
Eq.\ (\ref{e16}) above, if the integration limits are chosen to match
the threshold for star detection and clipping.  We find a very good
correlation between the deviation of the fit and the value of the
residual fluctuations in the patch; if all the histograms had the same
non-Gaussian shape the two should be proportional. Hence, the lower
limit radius can be approximated to depend on $\sigma$ itself via
\begin{equation}
R_i =[L_i/(4\pi N_{\rm cut} \sigma)]^{1/2}.
\label{e17}
\end{equation}
Equations (\ref{e16}), (\ref{e17}) then allow us to determine
$\sigma(b)$ for a given distribution of stars.

We now derive the latitude dependence implied by these equations.  As an
example, we use a plane parallel exponential model for stars in the
Galaxy: $n_i = n_{0,i} \exp(-\frac{r}{h_ix})$. This should be valid if
our clipping radius $R$ is small compared to the exponential scale
length for the Galactic disk $\alpha^{-1} \simeq $ (3 - 4) Kpc, and  is
in rough agreement with the disk in hydrostatic equilibrium with the
Galactic gravity field (Mihalas \& Binney 1981). Now Eq.\ (\ref{e16})
becomes
\begin{equation}
\sigma^2(x) = \sum_i R_i^{-1} \frac{ L_i^2 n_{0,i}\omega}{4\pi} E_2(\frac{
R_i }{ h_i x }),
\label{e18}
\end{equation}
where $E_n(u)\equiv \int_1^\infty t^{-n} \exp(-ut) dt$. We express
(\ref{e18}) in terms of the total number of stars in the $dR_i$
interval:
\begin{equation}
\frac{dN_i}{dR_i} = n_{0,i} \omega R_i^2 \exp(-\frac{ R_i }{ h_i x }).
\label{e19}
\end{equation}
Then the equation for $\sigma(x)$ becomes:
\begin{equation}
1=N_{\rm cut}^2 \sum_i
\left(R_i \frac{ dN_i }{ dR_i }\right)
[\exp(q)E_2(q)]|_{q= { R_i }/{ h_ix } }.
\label{e20}
\end{equation}
Consider first the limit when $R_i \ll h_i$. In this case $q<1$ and the
term in square brackets  is near unity. The number of clipped stars per
beam-width then becomes $N_> = \omega \sum_i \int_0^{R_i} n(r;x) r^2 dr
\simeq 3^{-1}  \sum_i (R_i dN_i/dR_i)$, leading to $N_> =N_{\rm
cut}^{-2}/3$. This is much smaller than we find in the data. Our
clipping algorithm removes 9 pixels per star and cuts deeper than
\Ncut=3.5 would indicate, and leaves only about 40\% of the pixels.  This
proves that  $q>1$, i.e.\ bright stars are seen out to  the scale height
of the disk and beyond.

We now consider the variations in star counts with magnitude in a
plane-parallel Galaxy. The differential counts in the magnitude interval
$dm$ are given by $dN/dm = \omega [d\ln r_m/dm] n(r_m) r_m^3$, where
$r_m \propto 10^{0.2m}$ is the radial distance to the star of apparent
magnitude $m$. In the limit of $r_m < h$, or $n\simeq$ constant, the
counts converge to the uniform distribution limit, $dN/dm \propto
10^{0.6m}$. In a plane-parallel Galaxy in which the radial structure can
be neglected, the number density of stars depends only on $|Z| \propto
10^{0.2m} \sin |b|$, and the  differential counts in the direction
$x={\rm cosec}|b|$ can be related to those at the Galactic pole by
\begin{equation}
\frac{dN}{dm}|_x = x^3 \frac{ dN(m-5\log_{10} x)}{dm}|_{\rm Pole}.
\label{e21}
\end{equation}
Hence, once the counts at the pole are measured, we can evaluate the
expected number counts  in any direction $b$ and then compute the
fluctuations in the flux they produce via
\begin{equation}
\sigma^2(x)  \propto \int_m^\infty 10^{-0.8 m} \frac{dN}{dm}|_x dm.
\label{e22}
\end{equation}
This formulation has two advantages: 1) the  data on star counts in the
polar regions can be used to evaluate the dependence of star
fluctuations on cosec$|b|$; and 2) the only assumption is plane
parallelism. We tested this assumption with both models and
observations.

We now determine $dN/dm$ from the DIRBE data, and compare with prior
observations. Figure \ref{f7} shows a Band 2 histogram of
$\log_{10}(|F-\langle F \rangle|)$ for a 128$^2$ pixel (38.4\deg
$\times$  38.4\deg) region at the North Galactic Pole (NGP). This
logarithmic form is useful for displaying both large and small
deviations, and we have normalized it to the equivalent number of stars
per magnitude per square degree.  The equivalent K magnitude is plotted
on the top horizontal axis. With the DIRBE beam of $1.42\times 10^{-4}$
sr (4.5 pixels),  $F =1$ \nwm2sr corresponds to $m_K$ = 9.45. Because
the CIB level and the smooth foreground contributions from dust and
faint stars are not known in advance, we must subtract the mean of this
confusion noise before seeking to measure brightnesses of individual
bright stars.   We define $\langle F \rangle$ as the mean flux of the
map clipped with \Ncut = 3.5; in this case it is  66 \nwm2sr or $m_K=5$,
near the turnover of the distribution. The  total number of pixels with
positive values of the flux after the subtraction is 12,359; the
remaining 24.6\% of the pixels have negative fluxes. We have plotted
both positive and negative deviations to illustrate the asymmetry of the
histogram far from the peak, as well as its symmetry near the peak.
Negative deviations are statistical fluctuations, while very bright
pixels are individual cataloged stars.   For the clipped map at \Ncut=
3.5, a total of 6,380 pixels remain. Out of these, 2,982 pixels (47\%)
have positive fluxes, showing the symmetry of the flux distribution for
the clipped map.

Figure \ref{f7} also shows a Gaussian distribution with the variance
$\sigma\simeq10$ \nwm2sr computed  with \Ncut = 3.5,  and for $\sigma=9$
\nwm2sr . The difference between these two Gaussians is not large on
this plot, showing the difficulty of direct detection of a
sub-population of Gaussian fluctuations with dispersion of 5 \nwm2sr .

The NGP star counts were observed directly by Elias (1978). We show his
data at  $K= 1$, 2.5, 3.25 and 8 with $N^{1/2}$ error bars and our
binning of his data. Further NGP data were obtained  by the 2MASS survey
in K$_s$ band, almost identical to the DIRBE Band 2, and were kindly
provided to us by Tom Jarrett (1998). The cumulative counts from these
measurements were shown in Fig.\ 1 of Beichman (1996) out to $K_s >15$,
who found that they follow $dN/dm \propto 10^{0.3 m}$  (cf.\  his Table
4). Actual 2MASS star counts from a region of 5 square degrees centered
on the NGP are plotted in Fig.\ \ref{f7}. The  agreement between the
DIRBE counts, the Elias (1978) and Jarrett (1998) data, and the $B=0.3$
extrapolation is excellent over 15 magnitudes, or six decades in flux.
South Galactic Pole counts  from Minezaki et al.\ (1998) are also shown,
and confirm the slope as well as the north-south symmetry of the Galaxy.
At $m_K < 1.5$, the counts tend to the slope of $B=0.6$ coming from
stars much closer than the scale height; if $B$ were less than 0.6, the
integrated star brightness would diverge at the bright end.

The star counts agree with model predictions. Both Beichman (1996) in
his Fig.\ 1 and Minezaki et al. (1998) in their Fig.\ 1 show  that the
counts are fitted well by extensions of either the Bahcall \& Soneira
(1980) or  Wainscoat et al.\  (1992) models. An eyeball fit to their
data gives $B=0.3-0.32$ at $K=11$.  The Wainscoat et al.\ model at
$K=11$ shown in Fig.\ 1 of Minezaki et al.\ (1998) gives $\log dN/dm
\simeq 1.35$, whereas continuation of the solid line in our Fig.\
\ref{f7} to $K = 11$ gives $\log dN/dm = 1.3$ if $B=0.3$ and 1.4 if
$B=0.33$. The agreement between the two slopes and normalizations is
thus very good. Even the large-beam DIRBE instrument sees far beyond the
scale height of the bright K band stars.

We can now estimate the star brightness fluctuations.  We begin with a
power law for star counts in a plane-parallel Galaxy, $dN/dm|_{\rm Pole}
\propto 10^{Bm}$. Then the fluctuation due to stars fainter than $m$ at
the Galactic pole becomes:
\begin{equation}
\sigma^2|_{\rm Pole} \propto \int_m^\infty 10^{-0.8m^\prime +Bm^\prime}
dm^\prime .
\label{e23}
\end{equation}
We  now derive the latitude scaling law implied by these assumptions. In
our clipping method the lower limit on $m$ in the integral is given by
$10^{-0.4 m} \propto N_{\rm cut} \sigma$.  Combining this with Eq.\
(\ref{e21}) we find that the fluctuation in the total star flux at
latitude $b$ is given by the power law
\begin{equation}
\sigma \propto (\cosec|b|)^{{1.2\over B}-2}.
\label{e24}
\end{equation}
A uniform, infinite star cloud has $B=0.6$, so there is no latitude
dependence in that case.  If $B=0.4$, then we  expect $\sigma \propto
\cosec|b|$, while if $B=0.3$ then $\sigma \propto \cosec^2 |b|$.  The
dependence of $C(0)$ on cosec$|b|$ can also be inverted via Eqs.\
(\ref{e21}), (\ref{e22}) to give a model for $dN/dm$. There is a unique
relation between the two functions, for a plane-parallel Galaxy and our
method of bright star extraction.

The value of \Ncut enters indirectly, since it determines the magnitude
$m$ of the cutoff of detected stars and hence the magnitude at which $B$
is important. Another measure of the star distribution function is the
number of  clipped stars, $N_>$. In the power law case we have a very
simple result:
\begin{equation}
N_> = {{0.8-B}\over B N_{\rm cut}^2}= \frac{2p+1}{3N_{\rm cut}^2},
\label{e25}
\end{equation}
where $p =  (1.2/B)-2$ is the logarithmic slope in Eq.\ (\ref{e24}). The
dependence of $N_> \propto N_{\rm cut}^{-2}$ agrees with our results
from DIRBE. Evaluating (\ref{e25}) for $B=0.6$ (or $r_m <h$) we find
$N_>=1/3N_{\rm cut}^2$, in agreement with the exponential distribution
of Eq.\ (\ref{e20}) with $h=\infty$, while for $B=0.3$ we find
$N_>=5/3N_{\rm cut}^2$.   The fraction of pixels clipped in the near-IR
DIRBE maps at \Ncut = 3.5 is $\sim 60\%$. Because the clipping algorithm
uses a mask with about twice the beam area, this corresponds to $N_> =
0.35$ so the effective \Ncut = 2.2 to 2.4  for $p=$ 2.0 to 2.5. The
effective \Ncut is less than the input value of 3.5 because each star is
observed in many pixels, each with its own noise.

Similarly, Eqs. (\ref{e21}), (\ref{e22}) allow us to evaluate the
expected variation of the residual  fluctuation with \Ncut,   $\sigma
\propto N_{\rm cut}^{-1 + 0.8/B}$. In the limit of a spatially uniform
distribution, $B=0.6$, and $\sigma \propto N_{\rm cut}^{1/3}$. For
$B=0.3$ it follows that $\sigma \propto N_{\rm cut}^{1.7}$, and reducing \Ncut 
from 7 to
3.5 should decrease $C(0)$ by a factor of 10, in  agreement with the
DIRBE data.

To summarize this section, we have shown that, for  $dN/dm \propto
10^{Bm}$ with $B=0.3$ out to $K<15$, and a plane-parallel Galaxy, we
should recover a unique dependence for Galactic star fluctuations,
$\sqrt{C(0)} \propto ({\rm cosec}|b|)^2$. This value of $B$ is
consistent with the \Ncut dependence of the fraction of clipped pixels
and the amplitude $\sigma$ of the remaining fluctuations in the DIRBE
data.

\subsection{Extrapolating to ${\rm cosec}|b|=0$} \label{ss44}

The scatter in Figs.\ \ref{f5} and \ref{f6} is  large, because the
plane-parallel model does not describe the large scale structure of the
Galaxy.  We divided the sample of 384 patches into four latitude bins
and plot them against Galactic longitude $l$ in Figs.\ \ref{f8} and
\ref{f9}. The dependence on  longitude is significant for latitudes even
as high as $|b| \sim 60\deg\ $.  However, for Galactic  longitudes
between 90\deg\ and 270\deg\ and high latitudes there is almost no
longitude dependence. We therefore selected data with $90\deg \leq l
\leq 270\deg $, and plot the dependence of $\sqrt{C(0)}$ on $\cosec |b|$
in Figs.\ \ref{f10} and \ref{f11}.  The scatter is much reduced.

We tried several fitting functions for the 127 patches for which
$|b|>20\deg$, $90\deg < l < 270 \deg$:
\begin{equation}
\sqrt{C_{\rm fit}(0)} = a + A ({\rm cosec}|b|)^{p},
\label{e26}
\end{equation}
where $a$ presumably contains the cosmological and instrumental parts of
the signal,
\begin{equation}
C_{\rm fit}(0) = a^2 +A^2 ({\rm cosec}|b|)^{2p},
\label{e27}
\end{equation}
\begin{equation}
\sqrt{C_{\rm fit}(0)} = a+\sum_{i=1}^2A_i({\rm cosec}|b|)^i ,
\label{e28}
\end{equation}
and
\begin{equation}
C_{\rm fit}(0) = a \exp(A {\rm cosec}|b|),
\label{e29}
\end{equation}
minimizing, for example, $ \langle [(C(0) - C_{\rm fit})/C(0)]^2\rangle$
with respect to $a, A, p$.   The fits for Eq.\ (\ref{e27}) are shown in
Figs.\ \ref{f10} and \ref{f11} and in Table 1. Except for the
color-subtracted $[2-3]$ maps, all fits give positive values of $a$. The
color-subtracted maps $[2-3]$ and $[3-4]$ have large scatter in the
color indices at which the variance $C_\Delta(0)$ is minimized,  and do
not allow for a robust determination of  $a$.  Note that $p>1$ as
expected from the model. The star and CIB fluctuations add in
quadrature, as we demonstrated by simulation, so functional forms such
as Eq.\ (\ref{e27}) are better justified than the other fits, although
all give consistent results. The last row in Table 1 summarizes our
limits on $a$ with errors corresponding to the extreme range from 92\%
confidence levels from all the fits, around our preferred central value
from Eq.\ (\ref{e27}).

We estimated the statistical uncertainty as follows. We define a
relative variance $\sigma_0^2\equiv \min \langle [({\rm data} - {\rm
fit})/{\rm data}]^2\rangle$ and a normalized $\chi_N^2(a, A, p) \equiv
\langle [({\rm data} - {\rm fit})/{\rm data}]^2\rangle/\sigma_0^2$.
Fig.\ \ref{f12} plots deviation histograms  vs.\ $(\delta_{\rm
fit}/\sigma)^2$ where $\delta_{\rm fit} \equiv (C(0) - a^2 - A^2 
x^{2p})/\sqrt{C(0)}$. For purely Gaussian deviations the
histograms would be straight lines of slope $-1/2$.

We plot contours
for $\Delta \chi^2 = 7$ which for 3 parameters $(a,p,A)$
corresponds to a confidence level of 92\%.
Fig.\ \ref{f13} shows thus determined confidence contours projected onto the 
$(a, 
p)$ plane for
Bands 1 to 4, according to Eq.\ (\ref{e27}).  The
uncertainties shown in  Table 1 correspond to the largest span of $a$ in
the panels in Fig.\ \ref{f12};  for any given value of $p$ the
corresponding uncertainty levels are reduced significantly. This is the
reason for the smaller formal uncertainties on $a$ when an exponential
fit - Eq.\ (\ref{e29}) - is assumed. Excluding from the analysis the
patches that lie close to the Ecliptic plane further reduces the scatter
in Figs.\ \ref{f10}, \ref{f11}, but  the contours are almost identical
to those in Fig.\ \ref{f13}. Likewise, keeping only the patches at $120
\deg < l < 240 \deg $ produces the same result.   The results for \Ncut
= 7 and 5 are consistent but with a larger uncertainty, whereas for
\Ncut = 3 the results are close to those shown in Table 1. The values
for $p$ are in good agreement with the expectations from the star
counts.  For Band 2, the maximal spread is $1.5 \leq p \leq 2.6$, which
corresponds to $0.26 \leq B \leq 0.34$ in good agreement with Fig.\
\ref{f7}.

To summarize, we have justified simple power law star count models and a
plane parallel model of Galactic fluctuations by comparing the
predictions to the measurements.  The model applies only far from the
Galactic center and Galactic plane. There is a statistically significant
residual term in the DIRBE data after extrapolation to zero cosec$|b|$,
with consistent values for four different fitting functions and for
different \Ncut. Plate 1 of our Paper II shows visually that there is no 
obvious
structure in the selected anti-center $|b|>20\deg$ data set.  We checked
the validity of our simple models by studying the residuals of the fits.
Fig.\ \ref{f14}  shows the residuals for Eq.\ (\ref{e27}) for each patch 
fluctuations, $\sqrt{C(0)}$  in Band 2.
There is no apparent correlation with Galactic longitude or latitude and the 
residuals are similarly independent of the ecliptic coordinates.

To consider whether the instrument itself or the zodiacal light could be
a source for this residual, we review the estimates in Paper I.  There we
constructed maps of the differences between different weeks of
observation, and found noise levels $\sqrt{C(0)}\sim$ 1.5, 0.3, 0.1
\nwm2sr in Bands 1 through 3 for \Ncut = 3.5. These are much smaller
than $a$ in Table 1, and it is unlikely that $a$ is due to noise.
Similar arguments apply to errors in the zodiacal light modeling, which
varies from week to week, and would contribute to the measurement error
calculation. In any case, in the near-IR bands the contribution of the
zodiacal light is small. It should be largest at 12 and 25 \um, where
the residual  fluctuations are below 1 \nwm2sr (Paper II and Sec.\
\ref{s5} of this paper), or $< 1 \%$ of the total foreground. Therefore,
the near-IR zodiacal modeling errors should be quite negligible compared
to $a$. Furthermore, zodiacal light has a sharp cusp near the ecliptic
plane, and zodiacal model errors would reflect this spatial dependence.
We tried excluding patches in or near the ecliptic plane and found no
change in $a,p$.  The spatial correlation function that we find for
\Ncut = 3 .5 (cf.\ Fig.\ \ref{f19}) is significantly different from that
expected for zodiacal light. The zodiacal light is smooth except near
the ecliptic plane, where dust resonances and asteroid family debris are
found.

We conclude that the single band plots all indicate a positive and 
approximately
isotropic residual
term that is unlikely to be produced by either instrumental noise or
errors in the zodiacal modeling. Since independent contributions add linearly 
to 
the combined variance, such a component would contribute only
$\sim10\%$ to the total dispersion of the confusion noise at the faint
end of the flux distribution plotted in Fig.\ \ref{f7}, and would not be
detectable there.  The Figure shows two Gaussians differing in
dispersion by 10\%, and they both seem reasonable fits.

\subsection{Analytic and Numerical Modeling} \label{ss45}

Since the measured residual $a$ can not be explained by known errors, we
must investigate the stellar foreground fluctuations more carefully. To
do so, we simulated the confusion noise process both analytically and
numerically.

Our analytical approach assumes a power law for $dN/dm$, with a Fourier
method to simulate the histogram of Fig.\ \ref{f7}. Let ${\cal P}(F)$ be
the probability distribution function to find a single star in the line
of sight with flux $F$.  The probability distribution function and its
characteristic function $G(f)$ are a pair of one dimensional Fourier
transforms, $G(f) = \int {\cal P}(F) \exp(ifF) dF$. The probability
distribution function of finding two stars is the convolution of ${\cal
P}$ with itself and the characteristic function for it is $G^2(f)$.
Similarly, for $n$ stars the characteristic function is $G^n(f)$. For
sufficiently large $n$ the characteristic function tends to a Gaussian,
which can be seen from expanding $G(f) \simeq 1 - f^2 \langle
F^2\rangle$, (Chandrasekhar 1954). For a Poisson distribution with $m$
stars per pixel on the average, the probability to find $n$ stars in a
given pixel is ${\cal P}(n)=m^n \exp(-m)/n!$.  Then  the complex Fourier
transform of the Poisson distribution of stars with an average of $m$
stars per pixel is $\exp(m(G-1))$. For many stars and sufficiently small
$f$ or  large fluctuations $G(f) \simeq 1-\frac{1}{2}f^2 \langle F^2
\rangle$, and the previous expression converges to a Gaussian
distribution whose width $\propto m^{-1/2}$. This prescription was
implemented numerically assuming ${\cal P}(F) \propto 10^{-
2.5B\log_{10} F}$ with $B=0.3$. Convolution of the predicted star
histogram with Gaussian measurement noise and Gaussian cosmic background
fluctuations can be included by multiplying $G(f)$ by a Gaussian. The
effective beam size sets the maximum value of the apparent $dN/dm$ in
the confusion noise region.     We are able to reproduce Fig.\ \ref{f7}
very well, including the negative fluctuations  and the transition from
the confusion noise  region to direct star detection. The result is
robust in that the effective value of $B$ measured at the bright end of
the distribution is not altered by subtracting the mean value, and
clipping with  various \Ncut has no effect on the value of $B$ at the
bright end.  The apparent number of bright stars is about 10\% larger
than the input value, owing to the bias introduced by undetected stars
of medium brightness near the detected bright stars.

We also want to know whether there is any feature of our processing
algorithm that could produce a spurious residual fluctuation $a$.  For
this, we need a numerical simulation of the sky, including both stars
and possible cosmic terms.  We used the DIRBE star model of Arendt et
al.\ (1998), who implemented the 96-component star population model of
Wainscoat et al.\ (1992), including spatial distribution models for the
disk, bulge, halo, ring and arm populations. These were evaluated using
the K-band $dN/dm$ for the central  pixel of each patch, and a
2-dimensional uniform random number generator, and the stars out to
$K=20$ were placed at the centers of the pixels in the map. The flux
from each star was then distributed among 9 pixels according to the
measured or assumed beam shape, using delta-function, uniform, and
Gaussian beam profiles. If the correct beam profile is used, the
simulated maps processed with our algorithms match the DIRBE histograms
in each patch very well.

We also simulated CIB fluctuations, starting with Gaussian fluctuations
in Fourier space with  power spectrum $P(k)=k^{-1.3}$, as expected on
the smallest DIRBE scales, and then multiplied by the top-hat beam
window function determined in Sec.\ \ref{s3}. The resultant field was
then Fourier transformed to real space and normalized to the modeled
cosmic variance, $\sigma_{\rm sky}$.  We added these maps to the
simulated star fluctuation maps and examined the results.

As was discussed in Sec.\ \ref{ss42},  our clipping algorithm clips
effectively about $\sim 1\ \sigma$ below the nominal value of \Ncut ,
because each star is seen in multiple pixels, each with its own noise
fluctuations. Furthermore, according to Eq.\ (\ref{e25}), there should
be no variation with cosec$|b|$ in the number of stars clipped in each
patch if the star counts follow the same $B$ over the relevant range of
magnitudes. To test this we computed the effective $N_{\rm
cut,eff}=\sqrt{\frac{5}{3}N_>^{-1}}$ according to Eq.\ (\ref{e25})
assuming $B=0.3$. The number of clipped stars, $N_>$, is half the number
of clipped pixels since our mask has twice the beam area. The results
are plotted against cosec$|b|$ in Fig.\ \ref{f15} for both the real sky
and the model Galaxy. The data show that in the anti-center quadrants
outside the Galactic plane, there is no trend with cosec$|b|$, and the
effective values of \Ncut have very small dispersion. For the real sky,
the mean in this range is $N_{\rm cut,eff}=2.26$ and the dispersion is
$\sigma_{\rm cut} = [\langle N_{\rm cut,eff}^2 \rangle - \langle N_{\rm
cut,eff} \rangle^2]^{1/2}=0.06$.  For the model sky, the mean $N_{\rm
cut,eff}=2.27$, and the dispersion is 0.07. The effective \Ncut is
independent of the simulated beam properties.

We found a simple way to test the assumption of a plane parallel Galaxy.
As Eq.\ (\ref{e21}) shows, the quantity $x^{-3}({dN}/{dm})|_x$ plotted
against $m-5\log_{10} x$ should be independent of $x= \cosec|b|$ and be
equal to the counts at the Pole. We computed histograms for 14 patches
of $64 \times 64$ pixels at $|b|>20 \deg$, $|\beta_{\rm ecl}| > 30\deg$
and $90\deg < l < 270\deg$. The patches were clipped at \Ncut = 3.5 and
the average flux for the remaining pixels of each patch was subtracted
from the map. The resultant distribution of $x^{-3} dN/dm$ is plotted in
Fig.\ \ref{f16} for K band,  with Poisson error bars for those points
that contain at least 25 pixels (or 5.5 stars). The figure confirms that
the stars are distributed in a plane-parallel way for this data set,
with $B\simeq 0.3$. The plot includes lines for the model Galaxy, with
error bands. They agree with the observations within the errors except
at the bright end, where the model is slightly low.

We simulated skies with a cosmic term of $\sigma_{\rm sky} = 5$ \nwm2sr
for the same $64 \times 64$ patches.  The observations and the simulated
model are plotted in Fig.\ \ref{f17}, with fluxes measured in units of
$\sqrt{C(0)}$. If all the fluctuations are due to stars drawn from a
power law distribution of the same slope, then all of the patches should
follow the same line on this diagram. The DIRBE data and the synthetic
Galaxy model look nearly identical, and there is no  noticeable
difference between $\sigma_{\rm sky} =5$ \nwm2sr and 0.  The DIRBE data
and the models could be matched even better by adjusting the beam shape
for the synthetic maps. The simulated confusion noise does not affect
the amplitude and slope of the star counts at $m_K < 5.5$.

Using our simulated models we also tested the scaling of the amplitude
of the residual variance at given \Ncut as a function of the input
isotropic component $\sigma_{\rm  sky}$. This is important because our
clipping algorithm might remove both cosmic  and star fluctuations in a
complicated and possibly non-linear way. We find that even  at \Ncut =
3.5, the residual variance $C(0)$ is to good accuracy the sum of the
variances  from stars and the simulated CIB  contribution. This shows
that Eq.\ (\ref{e27}) is better justified than the other choices  for
extrapolating to cosec$|b|=0$. The choice of beam shape - Gaussian,
delta-function, or uniform - has no effect on the effective \Ncut, but
does lead to a systematic change in the residual $C(0)$, with the
largest beam area having the highest variance.

Similarly, changing the size $f_{\rm size}$ of the lower envelope region
used by the point source recognition algorithm has no affect on $N_{\rm
cut,eff}$, but leads to a systematic dependence of $C(0) \propto
1/f_{\rm size}$. Clearly, the larger the value of $f_{\rm size}$, the
more diligently the algorithm recognizes point sources.  We used $f_{\rm
size}=5$, corresponding to about 5 DIRBE beams.

We can now predict what the Galactic star counts ought to be in order to
reproduce the $C(0)$ vs.\ $x = {\rm cosec}|b|$ relation found in the
previous section. Combining Eqs.\ (\ref{e21}) and (\ref{e22}) shows that
for a plane-parallel model the star contribution to the fluctuation is
related to $dN_P/dm$ at the Pole via
\begin{equation}
x\sigma^2(x) = F_0^2 \int_{m_L - 5 \log_{10} x}^\infty
10^{-0.8y}\frac{dN_P}{dy} dy,
\label{e30}
\end{equation}
where $F_0$ is the zero magnitude flux; for Band 2 it is 6050 \nwm2sr
per DIRBE pixel. The lower magnitude for our clipping method is given by
$m_L = -2.5 \log_{10}[f_m N_{\rm cut}\sigma(x)/F_0]$. Here $f_m$ is a
factor accounting for the  beam and the lower envelope used; based on
the discussion in the previous paragraph we expect $f_m \sim 0.5-0.6$.
Differentiating both sides of Eq.\ (\ref{e30}) leads to
\begin{equation}
\frac{dN_P}{dm} = 10^{0.4m} \;\; \frac{0.4\ln10}{f_m N_{\rm cut} F_0} \;\;
\frac{\partial[x\sigma^2(x)]/\partial x}{\partial[x^2\sigma(x)]/\partial x}.
\label{e31}
\end{equation}
Here the right-hand-side should evaluated for $x$ given by:
\begin{equation}
m(x)=-2.5\log_{10}[f_m N_{\rm cut}x^2\sigma(x)/F_0].
\label{e32}
\end{equation}
Eqs.\ (\ref{e31}), (\ref{e32}) form a closed set to determine the star
counts required to reproduce the $\sigma(x)$. They thus provide an
important consistency check between the measured effective \Ncut , the
$C(0)$ - $x$ relation, and the star counts at the Galactic Poles. They
also show whether the cosec$|b|$-independent part of $C(0)$ can be
produced by the observed stars.

We can measure the $f_m$ factor by normalizing the recovered star counts
to the DIRBE data at $m_K=4$, where the confusion noise  is negligible,
and find $f_m \simeq 0.6$. The solid line in Fig.\ \ref{f7} shows the
recovered star counts at the Pole from $C(0)$ given by Eq.\ (\ref{e27})
with $a=0$ with parameters ($A,p$) taken from Table 1. The agreement
between the  star  counts inverted from the observed $C(0)$ - $x$
relation and the actual data is remarkably good,  considering the
statistical uncertainty in $A,a$ and $p$, and given that all three
($C(0)$ vs.\ $x$, \Ncut and $dN_P/dm$) were determined independently.
The dashed line in Fig.\ \ref{f7} shows the star counts required to
reproduce Eq.\ (\ref{e26}) with $a = 4.8$ \nwm2sr according to  Eqs.\
(\ref{e31}), (\ref{e32}). The line overshoots the data by several
standard deviations at $m_K >5$  and  shows that this value of $a$
cannot be produced by the observed Galactic stars.

We can now use the simulated sky maps to test the extrapolations to
$\cosec|b|=0$. We constructed simulated data for 384 patches of $32
\times 32$ pixels which contained both the   K Band  Galaxy synthetic
model, and a contribution from the CIB with $\sigma_{\rm sky}$ varying
from 0 to 20 \nwm2sr .  Processing the simulated maps with the standard
algorithm, we find that the model without the CIB term is a good match
to the DIRBE data at intermediate latitudes, but is significantly
steeper as cosec$|b| \rightarrow 0$ at $|b| > 45\deg$, and has a zero
intercept within the statistical errors. On the other hand the simulated
data curve at $|b| > 45\deg$ flattens out for positive values of
$\sigma_{\rm sky}$.

We also tried fitting the simulated sky fluctuations to the
observations, using $C(0)= A_m \sigma^2(\sigma_{\rm sky}) + a_m$, and
found the effective gain and offset $A_m,a_m$. This approach has the
advantage of including all that is known about the geometrical shape of
the Galaxy, since the simulated maps use the Wainscoat et al.\ (1992)
shapes for the disk, bulge, halo, ring and arm.  If the model  includes
a cosmic term of the correct amplitude, then we should find $A=1$,
$a_m=0$. We find that for a model $\sigma_{\rm sky}=0$ the value of $A_m
= 1.4\pm 0.3$, consistent with unity, and the numbers for $A_m$ for
positive values of $\sigma_{\rm sky}$ are similar.  We could achieve
$A=1$ by better modeling of the DIRBE beam size, since we have already
shown that relationship.

We also confirm the values for the residual fluctuations $a$ in Table 1.
If the modeled sky has $\sigma_{\rm sky}=0$, we find $\sqrt{a_m} =
7.3^{+2.0}_{-2.8}$ \nwm2sr in agreement with Table 1.  Conversely, if we
choose a model $\sigma_{\rm sky}$ similar to $a$ in Table 1, then $a_m$
should be consistent with zero.  We find $a_m = 0$ with the 92\%  confidence
level for 2.8 \nwm2sr $< \sigma_{\rm sky} <$ 8.3 \nwm2sr  with the central
value lying at $\sigma_{\rm sky}=5.2$ \nwm2sr .

We can imagine only one possible feature in the DIRBE data processing 
that could lead to a false conclusion about the latitude dependence. 
We have shown that for a range of fixed beam size and shape, the sky 
models all predict a dependence of star fluctuations on latitude that 
extrapolates to zero fluctuations at zero $\cosec|b|$. There is 
however some possibility that the effective beam size might depend on 
the line of sight and hence possibly on Galactic latitude.  To 
explain the measurements, we would have to find an effect that 
systematically changes the effective beam area by an amount of  the order of 
30\%, comparable to the fraction of the fluctuations at the Galactic 
pole that seem to be of cosmic origin.

The DIRBE team measured the actual beam size very carefully from 
transits of bright stars through the beam, and did not find a 
significant dependence of beam profile on time or direction. 
However, as noted elsewhere, the effective beam profile for measuring 
fluctuations depends on several other effects.  The effective beam 
area is approximately $\Omega = 2 \pi (\theta_{\rm beam}^2/2 + 
\theta_{\rm pixels}^2 + \theta_{\rm pointing}^2)$, where $\theta_{\rm beam}$ is 
the measured beam radius, $\theta_{\rm pixels}$ is the rms radius of the 
pixels, and $\theta_{\rm pointing}$ is the rms (vector) pointing error. 
The rms pixel size is $d_{\rm pix}/6^{1/2} = 0.132\deg$ for a square 
pixel of side $d_{\rm pix}$, but for a rhombus of the same area and a 
60\deg corner angle, it is increased by a factor of 
$(2/\sqrt{3})^{1/2} = 1.075$ to 0.142\deg.  Ignoring pointing error, 
this increases the effective beam area by 3.2\%, a negligible amount 
in this context.  The measured DIRBE pointing error is 1.5 arcmin (1 
$\sigma$), and increases the effective beam area by a fixed 3.9\%. 
If, however, the pointing error were much larger than indicated by 
the statistics of the residuals from the pointing solution fits, and 
in addition were strongly dependent on Galactic latitude, the effect 
could be important for us.   The pointing solution was the subject of 
extraordinary scrutiny by the COBE team, and such errors would have 
been noticed.  We conclude that the latitude fitting method is not 
subject to errors due to changes of the effective beam size with 
latitude.

\subsection{Power spectrum} \label{ss46}

Although the dominant spatial structure of the near IR maps is simply
the white noise of stars, it is interesting to see whether large scale
averages could reveal a CIB component. We computed both power spectra
and angular correlation functions, and the results are shown in Figs.\
\ref{f18} and \ref{f19}.  We describe the power spectra first.

We used 96 patches of 64$\times$64 pixels (or 19\deg$\times$19\deg ),
and computed spatial power spectra without star clipping, and with
$N_{\rm cut} =$ 7 and 5, leaving over 90\% and 80\% of the pixels
remaining at high latitudes. Smaller \Ncut left too few pixels for
reliable power spectra, and showed significant effects of the masking.
We computed power spectra for single bands and for the color-subtracted
maps, $\delta_1 - \beta \delta_2$, with $\beta$ evaluated for each patch
according to Eq.\ (\ref{e15}).  After each power spectrum was
calculated, it was divided by the beam window function discussed in
Sec.\ \ref{s3} to take out the instrument signature.  The fact that the
power spectra are approximately flat confirms that stars are the
dominant sources of fluctuations for \Ncut = 5, and that the effective
window function, accounting for pixelization, map distortion, beam
smearing, and pointing errors, is correct.

The single band power spectra also show large-scale gradients produced
by the Galactic structure. The amplitude of the power spectrum decreases
with $N_{\rm cut}$, but the overall shape does not change appreciably,
indicating that the beam mask effects are negligible for these values of
\Ncut . The amplitudes of the power spectrum for all angular scales are
too large, and the shape too wrong, to allow for detection of the CIB
structure. However, for some patches at  $N_{\rm cut}=5$ the spectrum is
close to that expected from Table 1. For \Ncut of 5 or more, we can not
extrapolate these power spectra to  cosec$|b| =0$ for most of the scales
probed by $P(q)$, and for smaller \Ncut there are too few pixels left.

Fig.\ \ref{f18} shows the power spectrum for  patch No.\ 7, at Galactic
$(l,b)=(115\deg, 61\deg)$, and ecliptic $(\beta_{\rm Ecl}, \lambda_{\rm
Ecl})$ = $(56\deg, 163\deg)$. In the single bands, the amplitude of the
power spectrum decreases strongly after point source removal, but the
shape remains approximately the same. The solid line in each panel shows
the CIB signal according to Table 1,  assuming that the CIB power
spectrum has $P \propto q^{-1.3}$. This is a valid approximation for
scales below one degree, but on large scales the power index of the CIB
power spectrum may be different. The power spectrum of the foreground
exceeds the estimated CIB by only a modest factor, particularly in Band
4. An instrument with a smaller beam might detect these fluctuations
directly.

Color subtraction significantly reduces the foreground structure, as
shown in Fig.\ \ref{f18}. Without clipping, we find a reduction by a
factor of $\sim 30$ between adjacent bands, although the spectra are
still flat from star fluctuation noise. Galactic stars are dominant in
Band 4 and zodiacal and cirrus emission are dominant in Band 5, so these
bands are not strongly correlated. For this patch, color subtraction of
Bands   4 and 5 shows structure dominated by cirrus dust, which has  $P
\propto k^n$ with a steep $n=-2$ to $-3$; see Sec.\ \ref{s5}. The solid
line in the lower panels shows the estimated small-scale CIB between
Band 1 and Band 2 with logarithmic slope of $-1.3$ and $C(0)$ taken from
Table 1. In Bands $[2-3]$,  we plot $\sqrt{C(0)}=$2 \nwm2sr .

For lower values of $N_{\rm cut}$ and $q^{-1} > 1\deg $, the amplitude
of $P(q)$ is comparable to that expected from the CIB according to Table
1. Furthermore, the slope of the power spectrum for this patch flattens
significantly in the color-subtracted maps, and at some scales and
bands  approaches the logarithmic slope expected from the CIB. However,
inspection of the power spectra shows that even in the color-subtracted
maps, we do not detect a cosmological signal for \Ncut $\geq 5$. The
spectra are all consistent with the white-noise stellar distribution and
large-scale Galactic gradients. To check for a cosmological power
spectrum consistent with Table 1 we would have to clip to much lower
levels of \Ncut where beam masking problems prevent reliable
determination of the power spectrum.

For \Ncut$<5$, it is better to compute angular correlation functions,
which are unbiased by masking effects, even for \Ncut = 3.5 where less
than 50\% of pixels remain. We evaluated $C(\theta)= \langle \delta
F({\mbox{\boldmath$x$}} +\theta) \delta F({\mbox{\boldmath$x$}})
\rangle$, averaging over the remaining pairs of pixels separated by
angular distance $\theta$. (The points at different $\theta$ are not 
statistically independent of each other. For \Ncut = 7 and 5 the correlation 
function is that of white noise, a delta function at zero lag, with a level
consistent with the power spectrum plotted in Fig.\ \ref{f18}. It
remains close to the zero-lag value on scales inside the beam, $\theta <
0.5^\deg$ and rapidly drops to very small (positive or negative) numbers
on larger scales. For \Ncut = 3.5 for patches with low values of $C(0)$
the correlation function flattens significantly, which would be
consistent with CIB structure at the levels of Table 1. Fig.\ \ref{f19}
shows the correlation function for Patch 7 at \Ncut = 3.5,  for 10 bins
of 1 degree width.  The straight line  illustrates the slope  of a CIB
correlation function $\propto \theta^{-0.7}$. The correlation function
amplitude and slope are consistent with Table 1 and a CIB
interpretation. Large-scale gradients in the star and dust populations
are  responsible for the positive correlation function at very large
angles.

A further comparison between the results in Table 1 and the power
spectrum analysis can be made in the following way. For \Ncut = 5 at
each angular scale $q^{-1}$, we evaluate the minimal value of $P(q)$ for
all the patches, and plot the value of a typical fluctuation,
$\sqrt{q^2P(q)/2\pi}$, versus $\theta= \pi/q$ in Fig.\ \ref{f20}. The
slope of the  data is roughly that of white noise, $\sqrt{q^2P(q)/2\pi}
\propto q$, implying that we are still seeing Galactic stars. The  upper
limits  over  angular scales of 2\deg\ $< \theta <$ 15\deg\ are:
\begin{equation}
\delta F_{rms}(\theta) \leq A (\frac{\theta}{5\deg})^{-1},
\label{e33}
\end{equation}
with $A$ = 6, 2.5, 0.8, 0.5 \nwm2sr for Bands 1 to 4 respectively. The process 
of finding an all-sky minimum of $P_2(q)$ produces a very smooth curve whose 
uncertainties are predominantly systematic. These
are strong constraints on galaxy evolution. The shaded areas represent
the power spectra according to Table 1, assuming that  the CIB $P(q)
\propto q^{-1.3}$.  Table 1 is consistent with the upper limits given by
Eq.\ (\ref{e33}), except in Band 4. In Band 4, Table 1 would require a
small scale for the turn-over in the CIB power spectrum, implying that
much of it comes from high redshifts.

\section{Results in mid- to far-IR Bands } \label{s5}

\subsection{Foregrounds and $C(0)$ analysis} \label{ss51}

At wavelengths greater than 10 \um, dust in the Solar system and the
Galaxy produce most of the foreground emission. These sources are smooth
on small scales, so do not necessarily  prevent  detection of CIB
fluctuations.  Odenwald, Newmark and Smoot (1998)  detected $< 100$
nearby galaxies in the DIRBE data at wavelengths greater than 10 \um,
but with the exception of M31 and  the Large and  Small Magellanic
clouds, the galaxies are unresolved. Because there are few discrete
sources in the mid and far-IR DIRBE data, we can clip the DIRBE maps to
lower values of $N_{\rm cut}$ and keep the same number of pixels as for
the near-IR bands.

Clipping to low values of $N_{\rm cut}$ would remove some fluctuations
in the CIB as well as the foreground.  If the CIB zero-lag signal were
as high as the upper limit found in the maps, clipping at $N_{\rm cut
}=2$, 2.5 or 3 levels would decrease the real $C(0)$ by  20\%, 10\% and
0.001\% respectively. Given that we find only upper limits, clipping
down to $N_{\rm cut}=2$ is safe but lower levels would require
interpretation.

Fig.\ \ref{f21} shows histograms of the number of pixels remaining in
each patch of 32$\times$32 pixels after removing point sources  with
$N_{\rm cut}=2$. Because of the extended nature of the foreground
emission in these bands, we removed large-scale gradients first, using
polynomials of order up to 4. Fig.\ \ref{f22} shows the variation of the
residual $C(0)$ with cosec$|b|$ at \Ncut = 2 for $90\deg < l < 270$\deg.
A significant fraction of the foreground emission still comes from the
zodiacal light, even after subtraction of the DIRBE zodiacal light
model, and contributes to the large scatter in the plots.  Fig.\
\ref{f22} also shows the dependence of $C(0)$ on Galactic latitude for
the 111 patches that also have Ecliptic latitude $|\beta_{\rm Ecl}| >
25\deg$. The scatter is reduced but not enough  to extrapolate to
cosec$|b|$=0.

In Bands 5 and 6 there is only a weak dependence on $b$, and the
amplitude of the typical fluctuations seen in these bands changes by
only $\sim 50\%$ between cosec$|b| = 1$ and 3, showing approximate
isotropy for $|b|\geq 20\deg$. We interpret the  lowest values of
$\sqrt{C(0)}$ as upper limits  on the CIB fluctuations.  They are slight
improvements over Paper II, and are shown in Table 2.

In Bands 7 and 8 the dependence on Galactic latitude is more prominent.
Extrapolation to cosec$|b|=0$ with Eq.\ (\ref{e26}) gives values of $a$
in agreement with the lowest $\sqrt{C(0)}$ shown in Table 2, but with
significant error bars. Therefore we again interpret the derived $C(0)$
as upper limits on the CIB fluctuations. The slope of $C(0)$ with
Galactic latitude is consistent with a plane parallel Galaxy
distribution, i.e.\ $\sqrt{C(0)} \propto {\rm cosec}|b|$. The
fluctuations are approximately proportional to the total brightness.

The color diagrams at these wavelengths are not as clean as in the
near-IR. In many patches no color correlations between the adjacent
bands exist, and the color indices show large variations across the sky.
Local variations in the parameters of the Galactic and solar system dust
(e.g. density and temperature) are expected on large angular scales.  No
improvement in the CIB limits was achieved with mid-IR color
subtraction.

\subsection{Mid- and Far-IR Power Spectra} \label{ss52}

We computed  power spectra for  the same 96 patches of 64$\times$64
pixels used for the near IR bands.   Both single band power spectra and
color-subtracted maps show a weak decrease with clipping threshold, a
consequence of the extended character of the foreground emission. The
reduction of the foreground with the color subtraction  is not as large
as in the near-IR bands. In order to preserve information about $P(q)$
at all angular scales, no gradients were removed before power spectrum
analysis.

The power spectra have a shape typical of the known cirrus distribution
(cf. Gautier et al. 1992).
The spectrum $P(q)$ is steep: $P(q) \propto q^{-n}$ with $n\simeq
(2.5-3).$ This is consistent with little small scale structure and,
hence, no strong dependence in the resultant $C(0)$ on $N_{\rm cut}$. It
is also consistent with the power spectrum of cirrus emission measured
on arc-minute and degree scales (Gautier et al. 1992; Wright 1998).
Fig.\  \ref{f23} shows power spectra for the same Patch 7 used in the
near-IR analysis. It shows the steep power spectrum typical of the
cirrus distribution with $P(q) \propto q^{-3}$ for most of the scales.
Note that some of the fluctuations at 12 and 25 \um\ can be due to
errors in modeling the zodiacal light.

As in the near-IR analysis, we evaluated the minimal values of $P(q)$
for each angular scale $q^{-1}$ across the entire sky. Fig.\ \ref{f23}
shows these minimal values of the fluctuation $\sqrt{q^2P(q)/2\pi}$  as
a function of angular scale $\pi/q$. The minimal values come close to,
but are not as small as, those in Table 2, which were evaluated at
\Ncut=2 after gradient subtraction. Even in the patches with the least
fluctuations, the power spectrum is still as steep as $P\propto q^{-2}$,
leaving the fluctuation $\sqrt{q^2P(q)/2\pi}\simeq$ const. This is
indicative of cirrus emission in all areas of the sky. Infrared sky
surveys with higher angular  resolution should be able to reduce this
contribution, and possibly uncover the CIB fluctuations.

\section{Conclusions} \label{s6}

\begin{figure}
\centering
\leavevmode
\epsfxsize=1.0
\columnwidth
\epsfbox{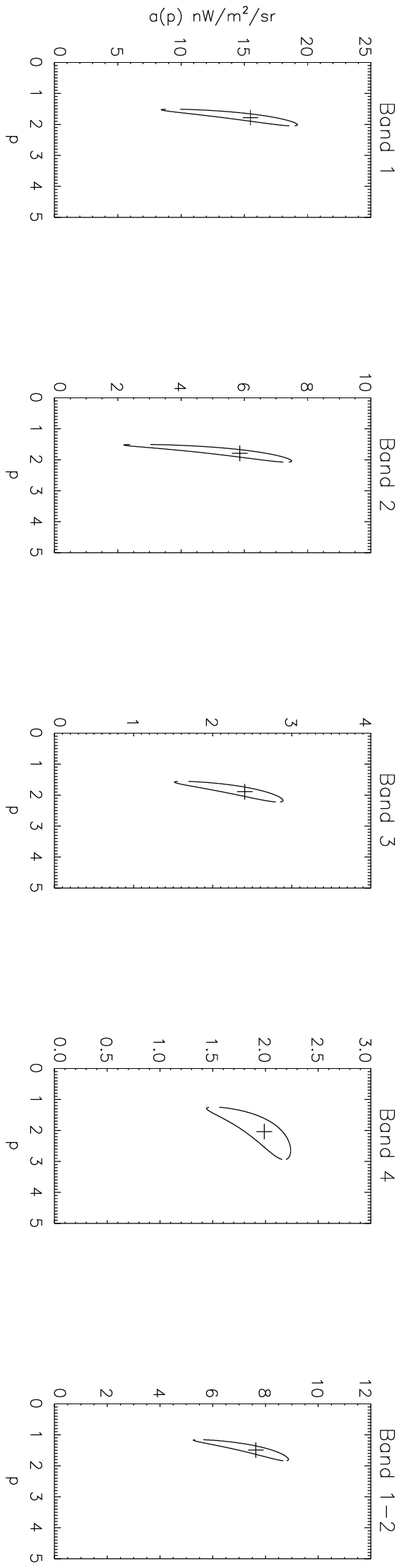}
\caption[]{ }
\label{f13}
\end{figure}

\begin{figure}
\centering
\leavevmode
\epsfxsize=1.0
\columnwidth
\epsfbox{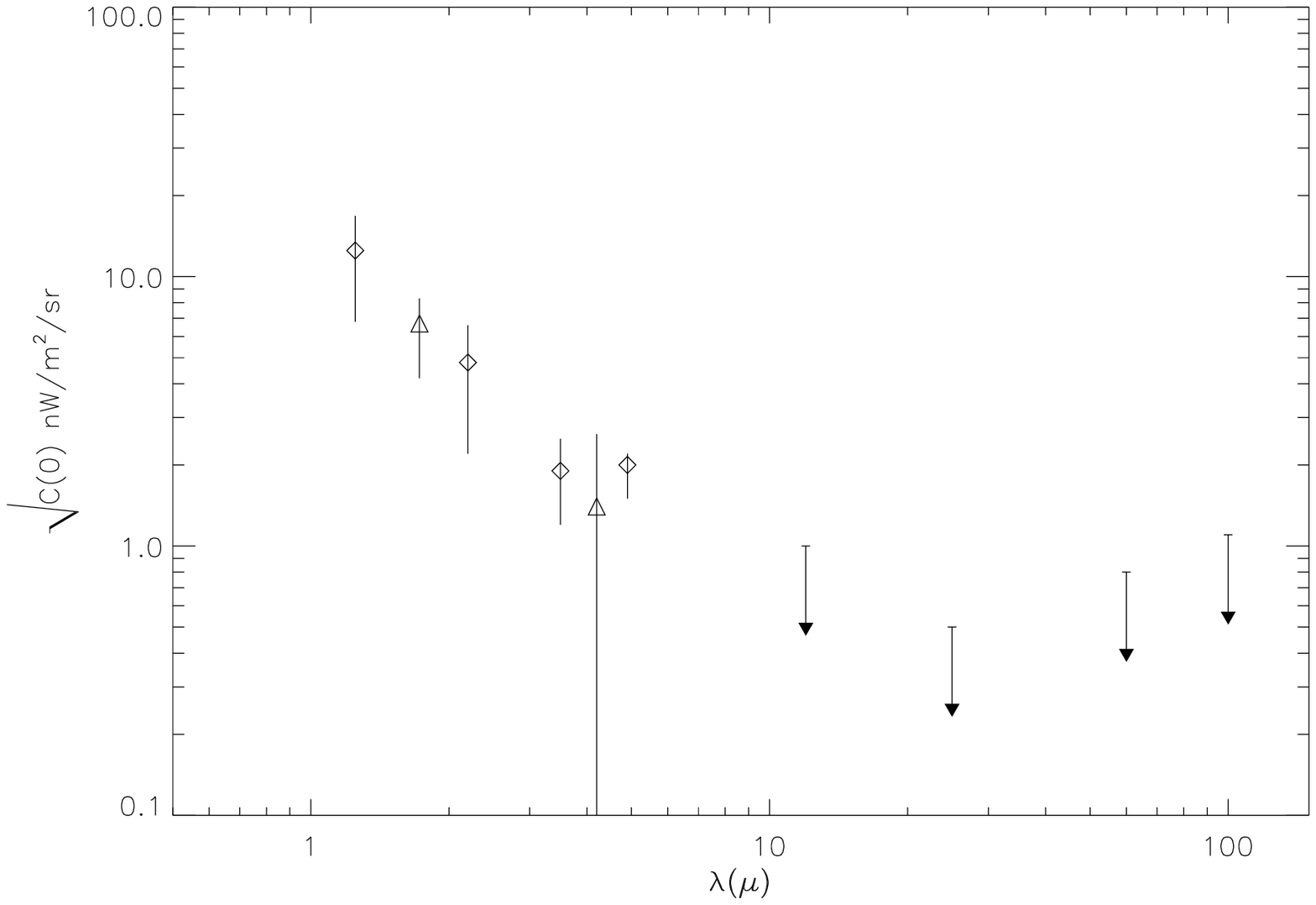}
\caption[]{ }
\label{f24}
\end{figure}

Fig.\ \ref{f24} summarizes our results. At  1-5 \um, we find a positive
residual fluctuation by extrapolating $C(0)$ to zero  $\cosec|b|$. Based
on detailed numerical and analytic models, this residual is not likely
to originate from the Galaxy, our clipping algorithm, or instrumental
noise.  We  conclude that this extra variance may result from structure
in the CIB. The variance found in  this way from individual DIRBE Bands
1-4 is plotted with diamonds, with  92\% uncertainty  levels from Fig.\
\ref{f13}. The results for color-subtracted maps are plotted with
triangles at wavelengths halfway between the bands used. We find a
positive residual in the color-subtracted map [1-2] between Bands 1 and
2, but not in other color subtracted maps. The color-subtracted map
[1-2] has a unique color of $\beta_{12}\simeq 2$ with little variation
across the sky and the limit measures $[\langle(\delta F_1 - 2 \delta
F_2)^2\rangle]^{1/2}$. Taken at face value, these high values of the near-IR 
CIB 
fluctuations, if produced by evolving normal galaxy populations, would require 
substantial CIB fluxes. These would have to be above the estimates from the 
K-band 
galaxy counts, but are below the upper limits found by Hauser et al. (1998). 
The 
upper limits on CIB fluctuations at  10 -
100 \um\ are plotted with arrows, and are below 1 \nwm2sr ,  with
$\sqrt{C(0)} < 0.5-0.7$ \nwm2sr at 25 \um. These limits are lower than
those in Paper II and imply strong constraints on how and when the early
galaxies formed and evolved.

On larger scales, $2^\deg < \theta < 15^\deg$, we obtain upper limits on
the CIB fluctuations from the all-sky power-spectrum analysis:
$(\theta/5^\deg)\times \delta F_{\rm rms}(\theta)  < $ 6, 2.5, 0.8, 0.5
\nwm2sr in Bands 1-4 respectively. These limits, when taken in
conjunction with our possible detection of the zero-lag CIB signal,
limit the turn-over scale in the spectrum of the primordial density
field to not much more than $\sim 100 h^{-1}$Mpc.

We fully recognize the difficulty of finding small fluctuations in 
the presence of larger fluctuations from foregrounds.  While we have 
found no local explanation for our results, it is still quite 
possible that the fluctuations are not of cosmic origin, but come 
from some fault of the instrument, the data processing, or an 
unexpected feature of the Galactic foreground. The best way to 
resolve this uncertainty is to get better data, such as from a higher 
resolution sky survey with exceptionally good attention to flat 
fielding.  This may be possible with satellites like SIRTF  or the 
proposed NGSS, or even from 2MASS data or rocket data.

We acknowledge support from the NASA Long Term Space Astrophysics grant
399-20-61-02. We particularly thank the COBE DIRBE team, led by the
principal investigator Michael Hauser and by Thomas Kelsall, for
developing an exceptionally well calibrated and stable instrument, and
for producing public data archives with the complex zodiacal light
foreground models removed.  We have benefited greatly from conversations
with Richard Arendt on applications of star models to the DIRBE data and
for a careful reading and constructive  comments on the manuscript of
this paper.  Our special thanks go to Tom Jarrett for providing the
2MASS star counts data. We thank the referee, Michael Vogeley, for careful 
reading of the paper.

\clearpage
\begin{table*}

\begin{tabular}
{c c c c c c c c c}
Fit & Band 1 & Band 2 & Band 3 & Band 4 & Band 1-2 & Band 2-3 & Band 3-4 \\
    &  1.25 \mic & 2.2 \mic & 3.5 \mic & 4.9 \mic & & & \\
\hline
$\sqrt{C(0)}=a+A x^p$
\\ $a$ & $12.5^{+4.3}_{-5.7}$ & $4.8^{+1.8}_{-2.4}$ & $1.9^{+0.6}_{-0.7}$ &
 $2.0^{+0.2}_{-0.2}$ & $6.7^{+1.6}_{-2.5}$ & $--$ & $1.4^{+1.2}_{-1.6}$ \\
$p$ & $2.15^{+0.41}_{-0.41}$ & $2.18^{+0.41}_{-0.41}$ &
$2.31^{+0.45}_{-0.45}$ & $2.79^{+0.61}_{-0.58}$ &
$1.92^{+0.55}_{-0.54}$ & $--$ & 1.47 \\
\hline
$C(0)=a^2+A^2 x^{2p}$
\\ $a$ & $15.5^{+3.7}_{-7.0}$ & $5.9^{+1.6}_{-3.7}$ & $2.4^{+0.5}_{-0.9}$ &
 $2.0^{+0.25}_{-0.5}$ & $7.6^{+1.2}_{-2.4}$ & $--$ & $$ \\
$p$ & $1.78^{+0.27}_{-0.27}$ & $1.79^{+0.28}_{-0.29}$ &
$1.89^{+0.33}_{-0.33}$ & $2.03^{+0.90}_{-0.78}$ &
$1.49^{+0.34}_{-0.33}$ & $--$ & 1.23 \\
\hline
$\sqrt{C(0)}=a+ \sum_{i=1}^2A_ix^i$\\
$a$ & 14.9$\pm5.4$ & 5.6$\pm2.2$ & 2.5$\pm1.2$ & 2.7$\pm1.2$ & $6.6\pm0.8$ &
 $--$ & $0.9\pm 0.9$\\
\hline
$\sqrt{C(0)}=a\exp(A x)$
\\ $a$ & $9.7\pm 0.6$ & $3.7\pm 0.2$ & $1.3\pm 0.1$ &
 $1.3\pm 0.1$ & $5.3\pm 0.2$ & $3.6 \pm 0.2$ & $1.5\pm 0.1$ \\
\hline
$x = [\ln\sqrt{C(0)}$-$\ln a]/A $
\\ $a$ & $8.8\pm 1.3$ & $3.3\pm 0.3$ & $1.1\pm 0.1$ &
 $0.8\pm 0.1$ & $4.8\pm 0.6$ & $1.8 \pm 0.4$ & $1.2\pm 0.1$ \\
\hline
Summary\\
\\ $a$ & $15.5^{+3.7}_{-7.0}$ & $5.9^{+1.6}_{-3.7}$ & $2.4^{+0.5}_{-0.9}$ &
 $2.0^{+0.25}_{-0.5}$ & $7.6^{+1.2}_{-2.4}$ & $--$ & $$ \\
\hline
\end{tabular}
Note: $x = {\rm cosec}|b|$
\caption{Limits on $a$ in \nwm2sr ; the range corresponds to 92\% confidence 
levels.
\label{t1}}
\end{table*}

\clearpage
\begin{table*}
\begin{tabular}
{c c c c c}
 & Band 5 & Band 6 & Band 7 & Band 8 \\
 & 12 \mic & 25 \mic & 60 \mic & 100 \mic \\
\hline
\\ $\sqrt{C(0)}$ & 1.0 & 0.5 & 0.8 & 1.1 \\
\hline
\end{tabular}
\caption{Upper limits on $[C(0)]^{1/2}$ in \nwm2sr for Bands   5-8.\label{t2}}
\end{table*}

\begin{figure}
\centering
\leavevmode
\epsfxsize=1.0
\columnwidth
z\epsfbox{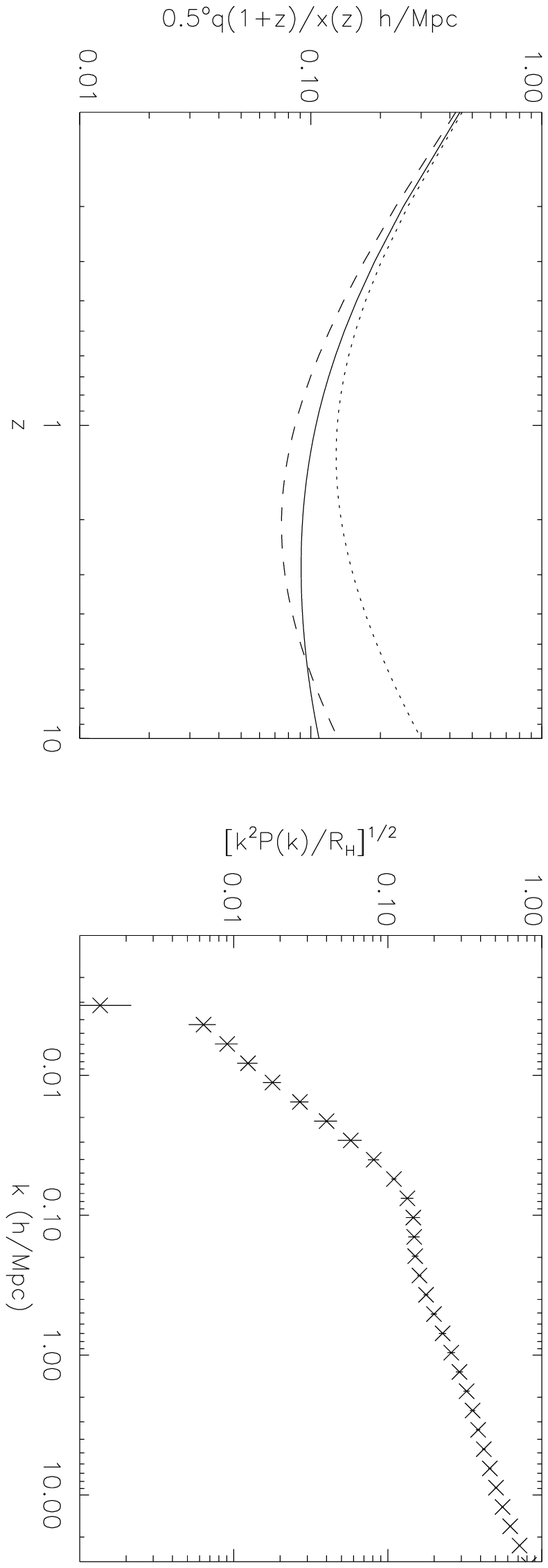}
\caption[]{ }
\label{f1}
\end{figure}

\clearpage
Fig.\ \ref{f1}: (a) Left: Linear scales  of 0.5\deg\ in Eq.\ (\ref{e9})
vs.\ z. The minimal scale is almost independent of cosmology.
(b) Right: $\Delta(k)  \equiv ({R_H^{-1}k^2 P_3(k; 0))}^{1/2}$ vs.\ $k$ for
APM spectrum. The
CIB minimal relative fluctuations on
the DIRBE beam scale are $\sim \Delta$ at the minimal value of the linear
scale.

\begin{figure}
\centering
\leavevmode
\epsfxsize=1.0
\columnwidth
\epsfbox{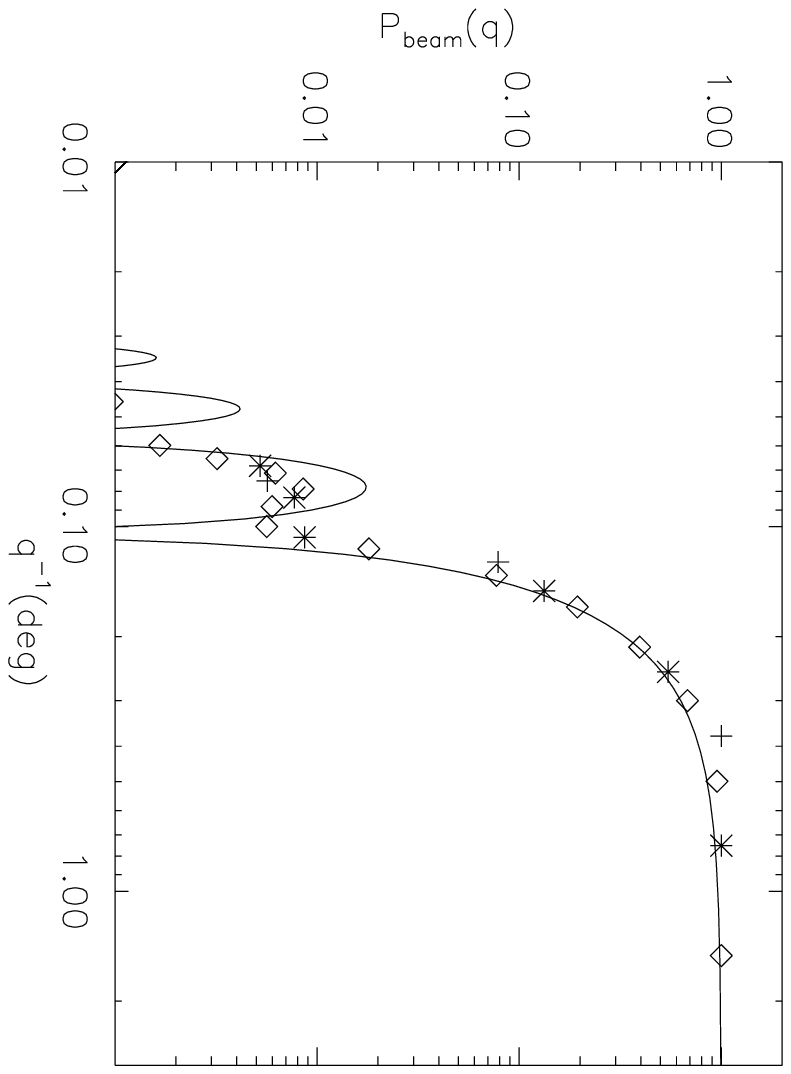}
\caption[]{ }
\label{f2}
\end{figure}

Fig.\ \ref{f2}: Window function and beam profile for DIRBE Band 1  beam.
Plus signs show
embedding in a 256$^2$ pixel field, asterisks 512$^2$ pixels, and diamonds
1024$^2$
pixels. Solid line
is a top-hat profile with a beam radius of
$\vartheta=0.4$\deg.

\begin{figure}
\centering
\leavevmode
\epsfxsize=1.0
\columnwidth
\epsfbox{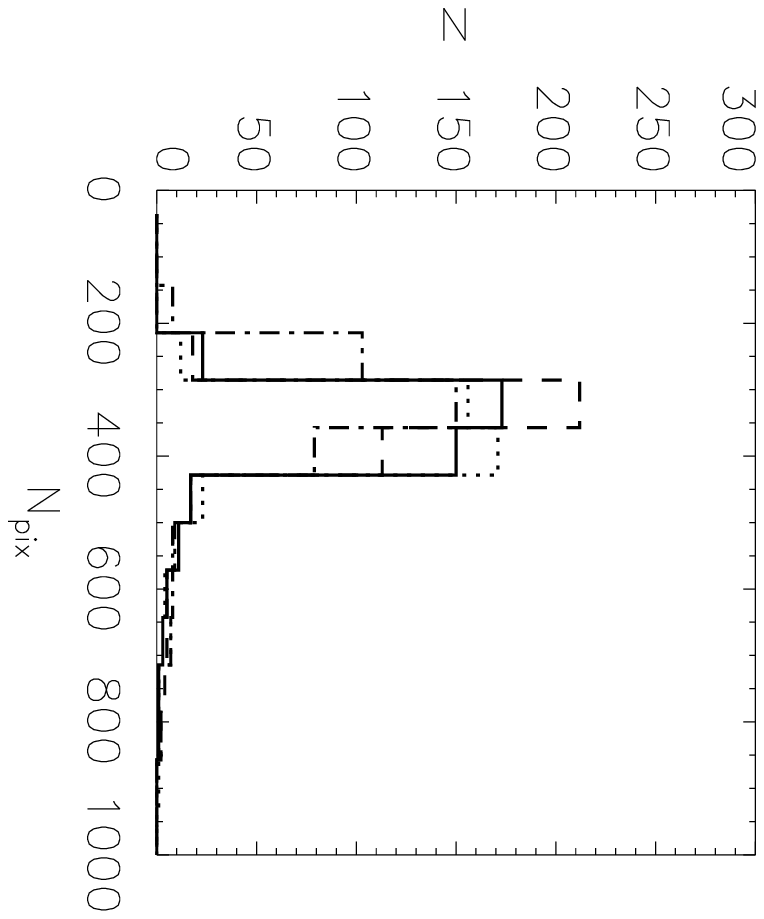}
\caption[]{ }
\label{f3}
\end{figure}

Fig.\ \ref{f3}: Histogram of pixels surviving clipping at \Ncut = 3.5 in
the near-IR
DIRBE bands for  384 patches of $32\times32$ pixels. Solid line is Band 1,
dotted  Band 2,
dashed  Band  3, and dashed-dotted  Band 4.

\begin{figure}
\centering
\leavevmode
\epsfxsize=1.0
\columnwidth
\epsfbox{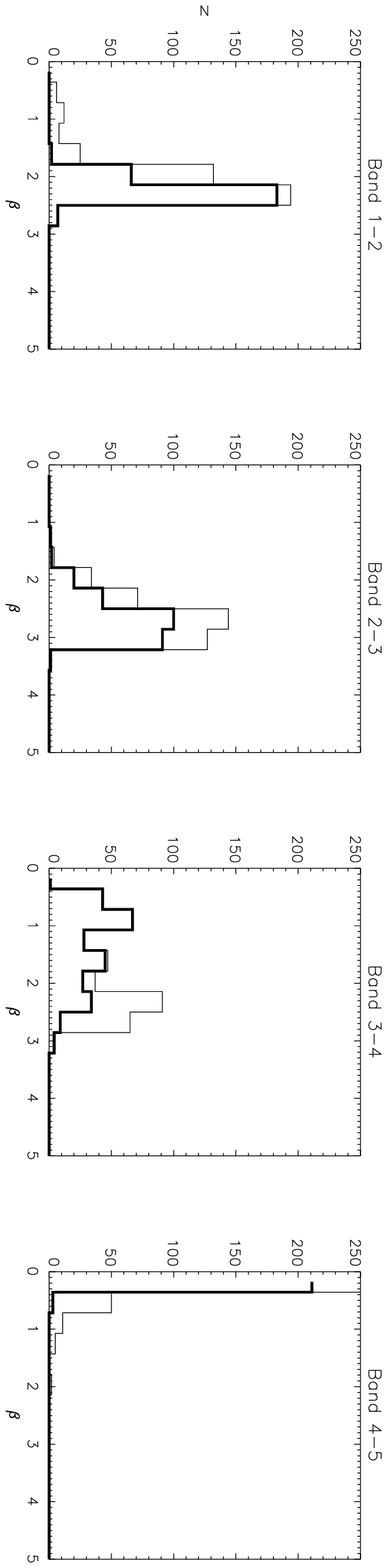}
\caption[]{ }
\label{f4}
\end{figure}

Fig.\ \ref{f4}: Histograms of color indices $\beta$ from Eq.\ (\ref{e15})
for 384 patches.
Thin solid lines are all-sky; thick lines are patches with $|b|>20$\deg.

\begin{figure}
\centering
\leavevmode
\epsfxsize=1.0
\columnwidth
\epsfbox{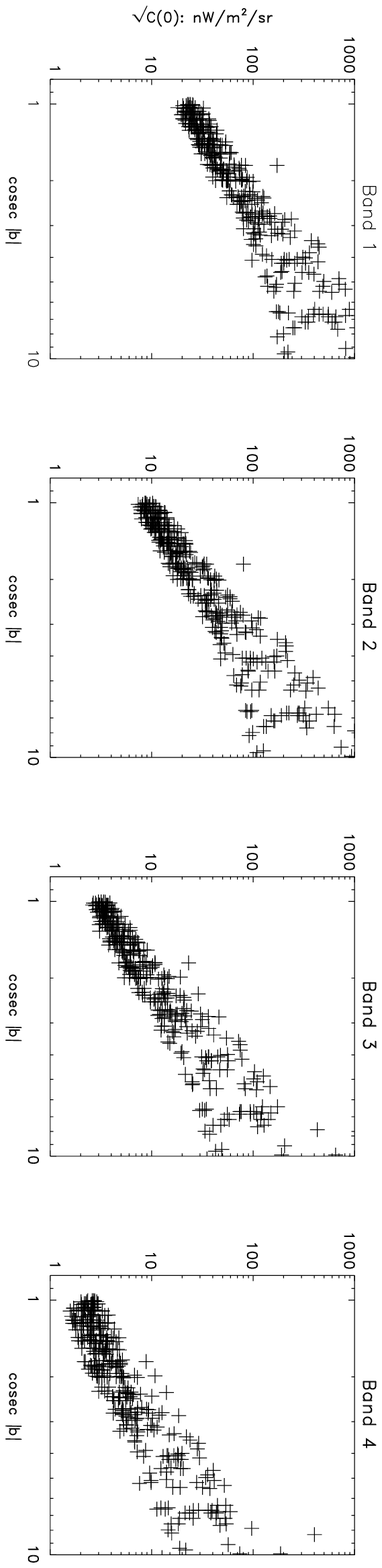}
\caption[]{ }
\label{f5}
\end{figure}

Fig.\ \ref{f5}: $\sqrt{C(0)}$ vs.\ cosec$|b|$ for Bands  1-4  and $N_{\rm
cut}=3.5$.

\begin{figure}
\centering
\leavevmode
\epsfxsize=1.0
\columnwidth
\epsfbox{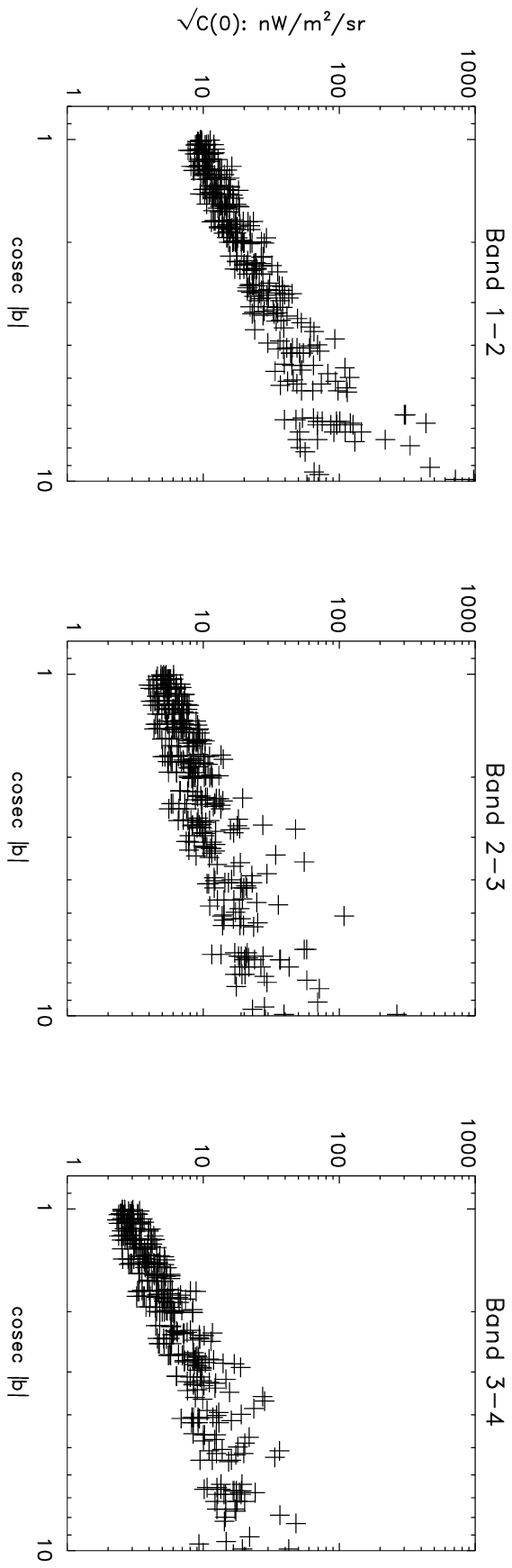}
\caption[]{ }
\label{f6}
\end{figure}

Fig.\ \ref{f6}: Same as Fig.\ \ref{f5} but for color subtracted maps.

\begin{figure}
\centering
\leavevmode
\epsfxsize=1.0
\columnwidth
\epsfbox{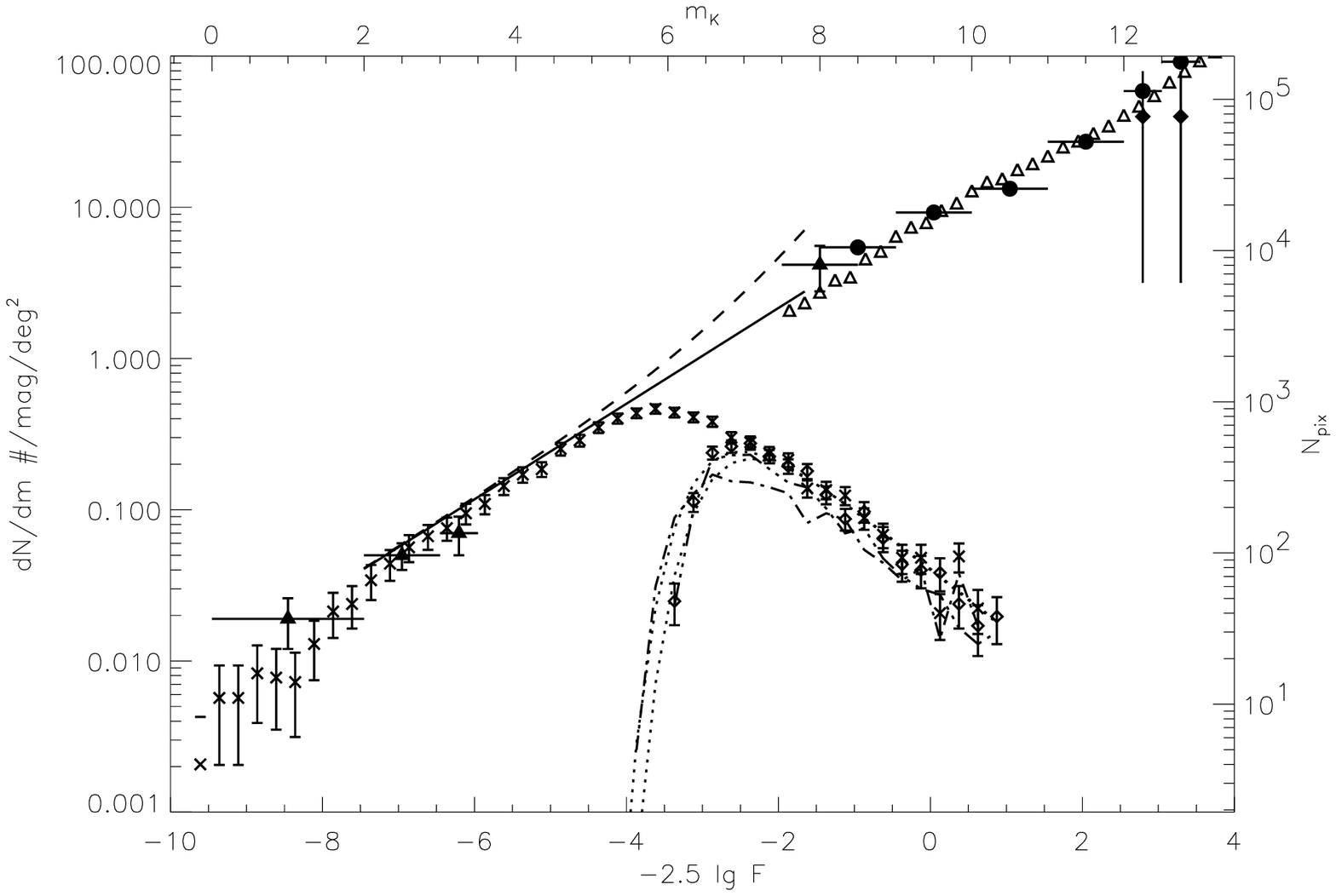}
\caption[]{ }
\label{f7}
\end{figure}

Fig.\ \ref{f7}: K band $dN/dm$ and DIRBE pixel histogram for 128$^2$ pixels
at NGP.  Flux
$F$ is absolute value of deviation from mean of patch after clipping with
\Ncut = 3.5,
measured in
\nwm2sr .  Positive deviations are + signs, diamonds negatives.  Poisson
error bars for
DIRBE data assume  4.5 DIRBE pixels per star.  Dotted lines are a Gaussian
fit to data after
clipping, and a 10\% lower dispersion. Dash-dots are positive pixels
remaining after
clipping and dash-dot-dot-dot shows remaining negative pixels.  Solid line
shows counts
inverted according to Eqs.\ (\ref{e31}) and (\ref{e32}) from Eq.\
(\ref{e27})  without
an isotropic component;
long dashes are inverted counts if  $a$ comes from stars.  Filled triangles are
differential counts  from  Elias (1978) NGP measurements. Filled circles
are differential
NGP counts from 2MASS  (Jarrett 1998). Open triangles are cumulative 2MASS
counts
multiplied by
$0.3\ln 10$ to convert to  differential counts for $dN/dm \propto
10^{0.3m}$. Filled
diamonds with error bars are South Galactic Pole counts from Fig.\ 1 of
Minezaki et al.
(1998).

\begin{figure}
\centering
\leavevmode
\epsfxsize=1.0
\columnwidth
\epsfbox{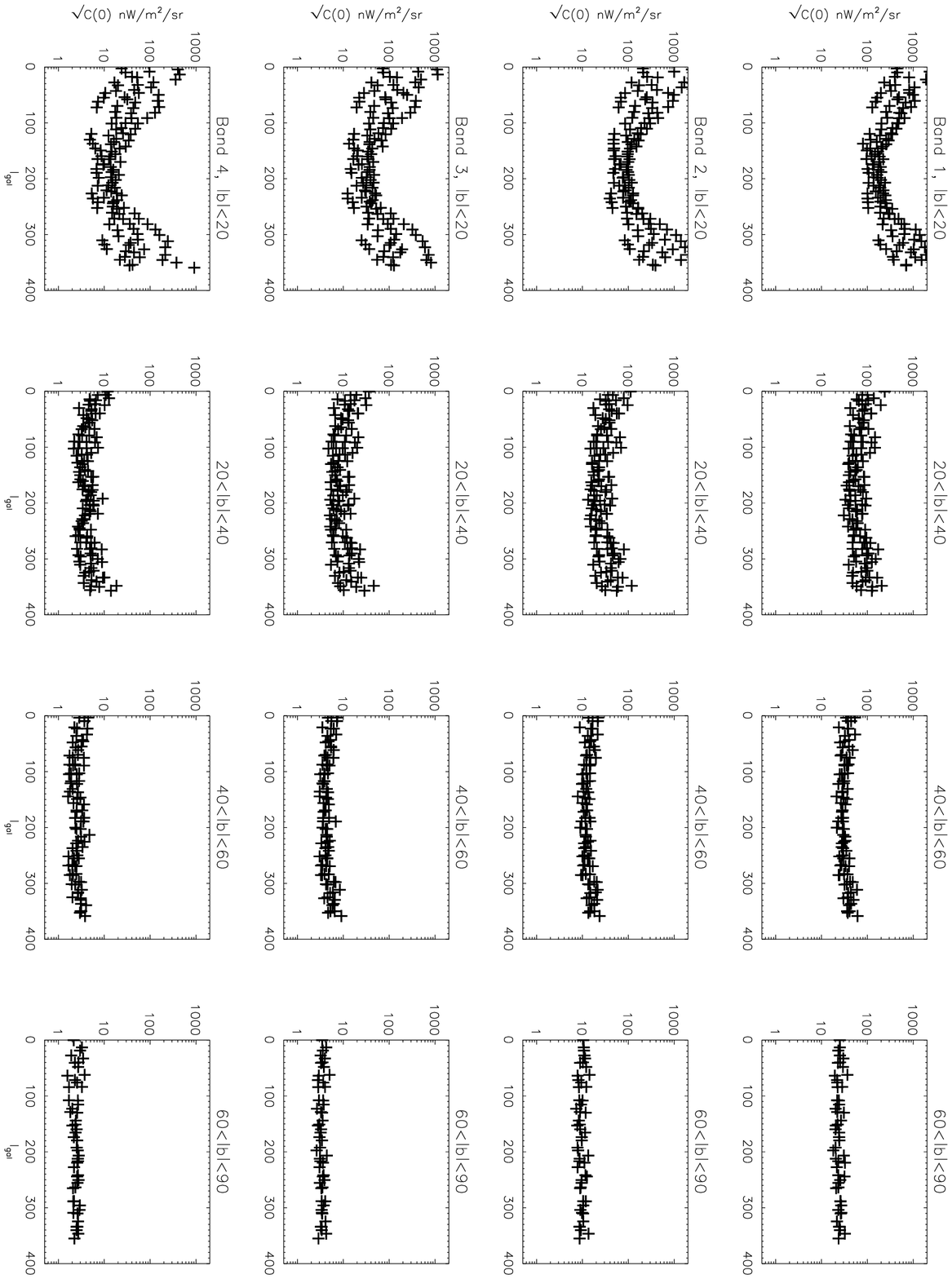}
\caption[]{ }
\label{f8}
\end{figure}

Fig.\ \ref{f8}: Longitude dependence of $\sqrt{C(0)}$ for individual DIRBE
bands
at various Galactic latitudes. The increase of fluctuations
towards the Galactic Center can be seen for any Galactic latitude.

\begin{figure}
\centering
\leavevmode
\epsfxsize=1.0
\columnwidth
\epsfbox{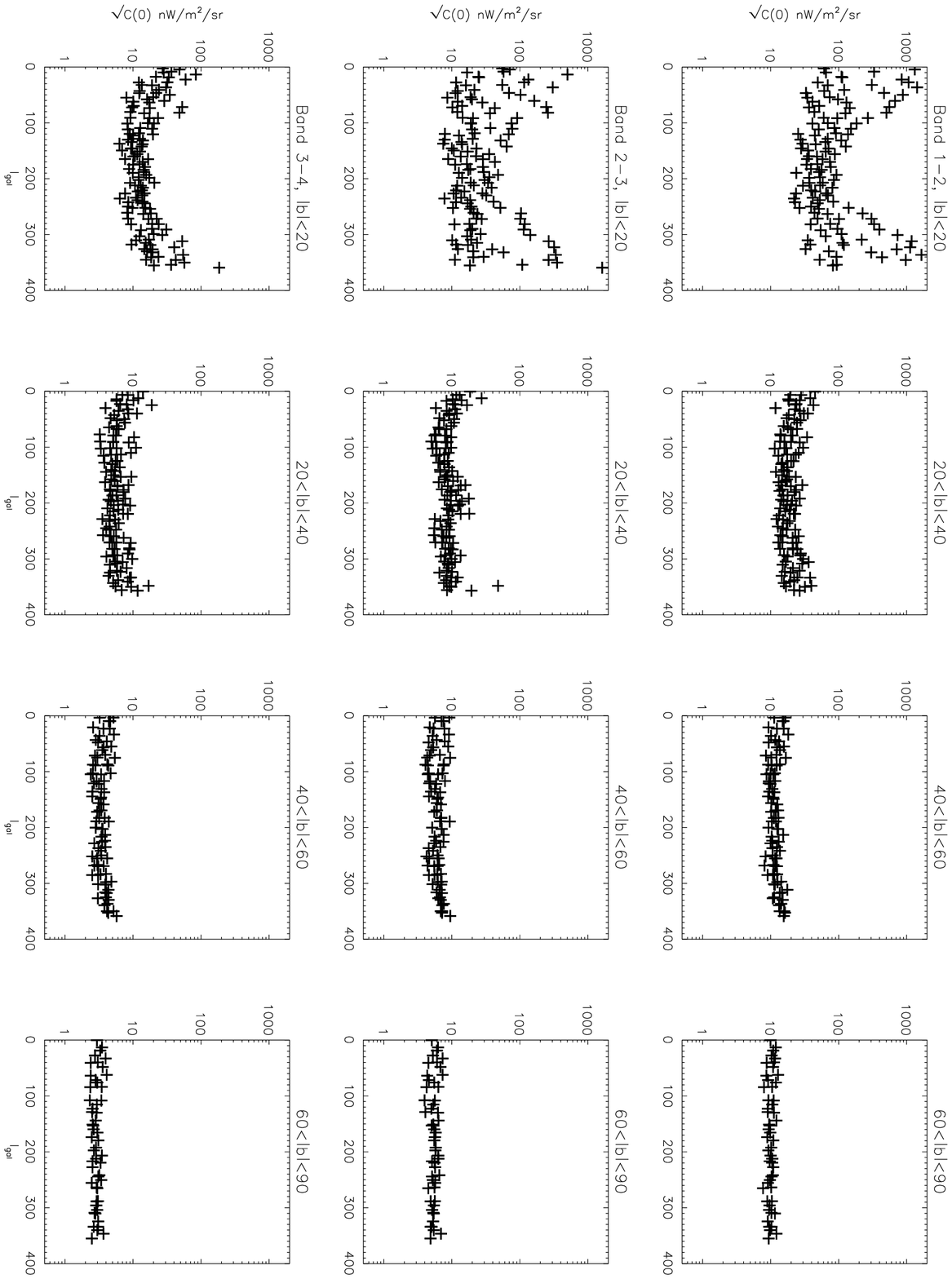}
\caption[]{ }
\label{f9}
\end{figure}

Fig.\ \ref{f9}: Same as Fig.\ \ref{f8} only for color-subtracted maps.

\begin{figure}
\centering
\leavevmode
\epsfxsize=1.0
\columnwidth
\epsfbox{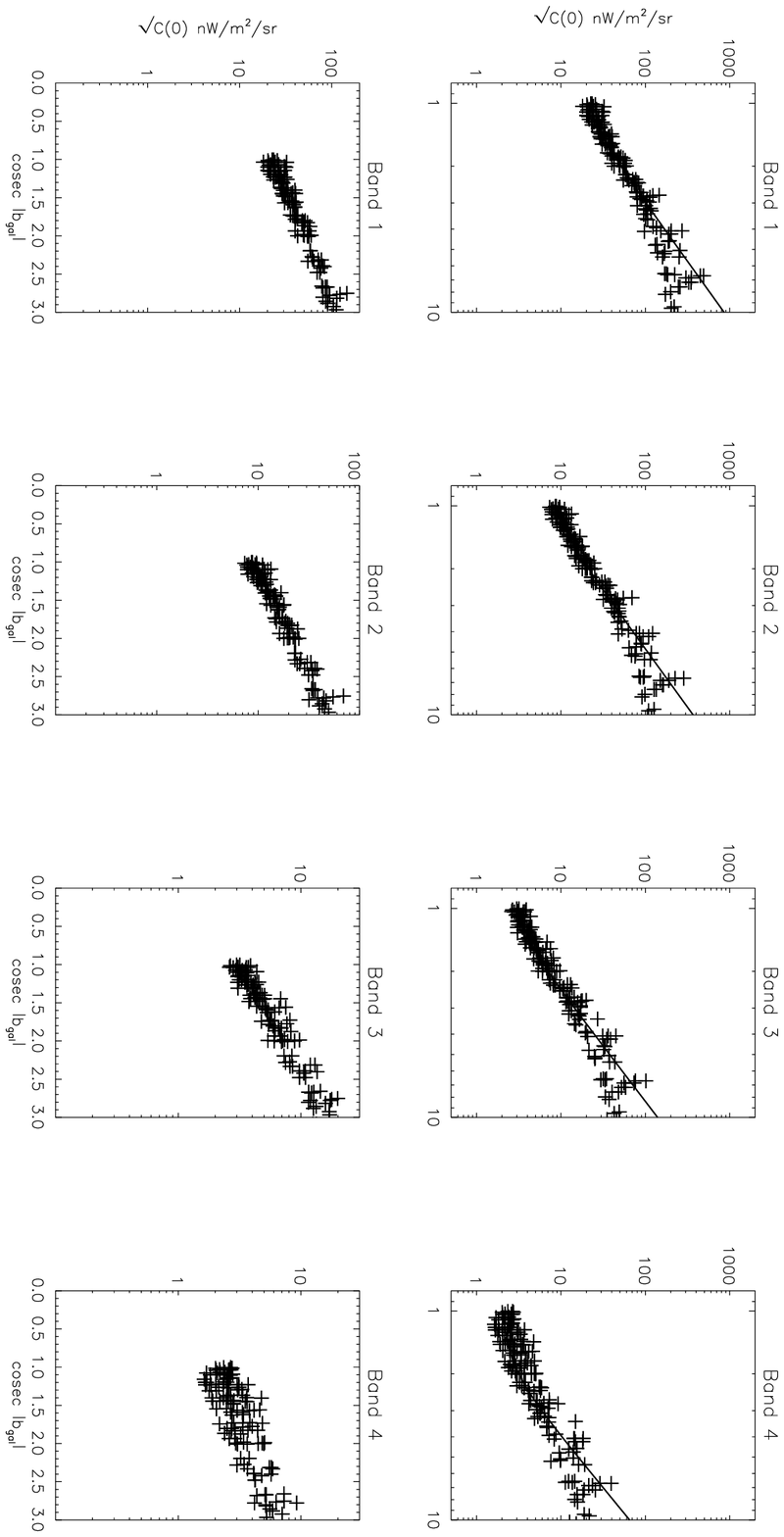}
\caption[]{ }
\label{f10}
\end{figure}

Fig.\ \ref{f10}: Plots of $\sqrt{C(0)}$ vs. cosec$|b|$ for J, K, L, M bands
for $90\deg < l < 270\deg$. Solid lines are fits of
Eq.\ (\ref{e27}) using data for $|b|>20 \deg$.

\begin{figure}
\centering
\leavevmode
\epsfxsize=1.0
\columnwidth
\epsfbox{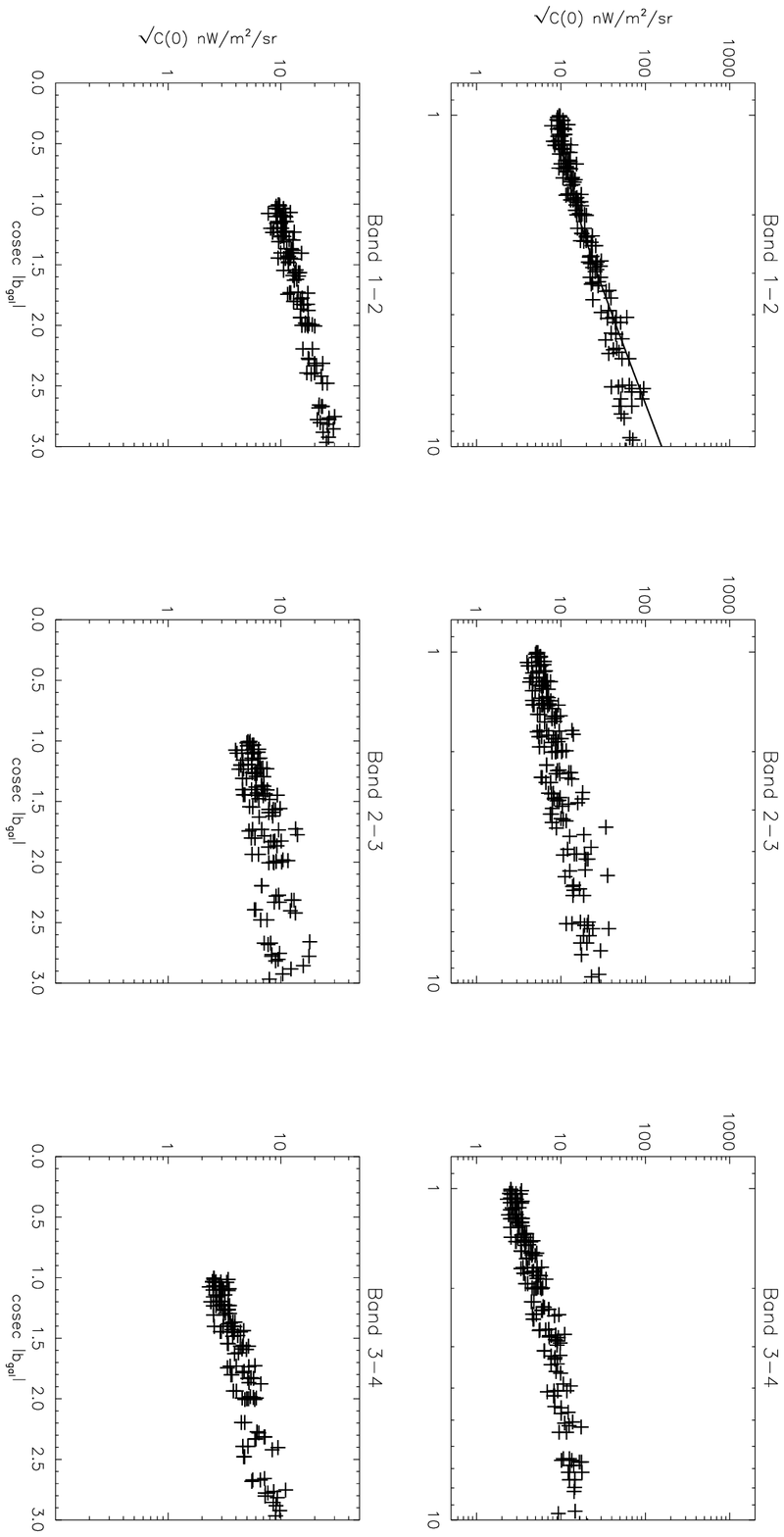}
\caption[]{ }
\label{f11}
\end{figure}

Fig.\ \ref{f11}: Same as Fig.\ \ref{f10} for color-subtracted maps.

\begin{figure}
\centering
\leavevmode
\epsfxsize=1.0
\columnwidth
\epsfbox{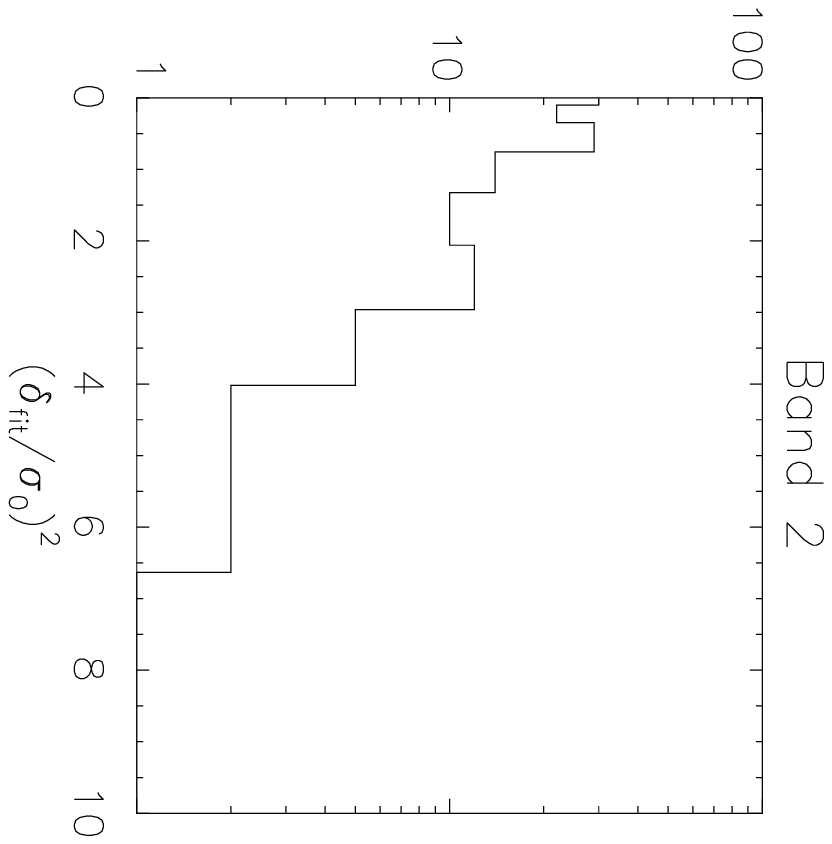}
\caption[]{ }
\label{f12}
\end{figure}

Fig.\ \ref{f12}: Histogram of fit residuals from Eq.\ (\ref{e26}) for Band 2
$\delta^2_{\rm fit}=  [(C(0)-a^2 -A^2 {\rm cosec}^{2p} |b|)/C(0)]^2$
in units of $\sigma_0$.

Fig.\ \ref{f13}: 92\% confidence limits on J, K, L, M and J-K for the fits
to Eq.\ (\ref{e27}). The plus sign  is the most likely value of $(a, p)$
from Eq.\ (\ref{e27}).

\begin{figure}
\centering
\leavevmode
\epsfxsize=1.0
\columnwidth
\epsfbox{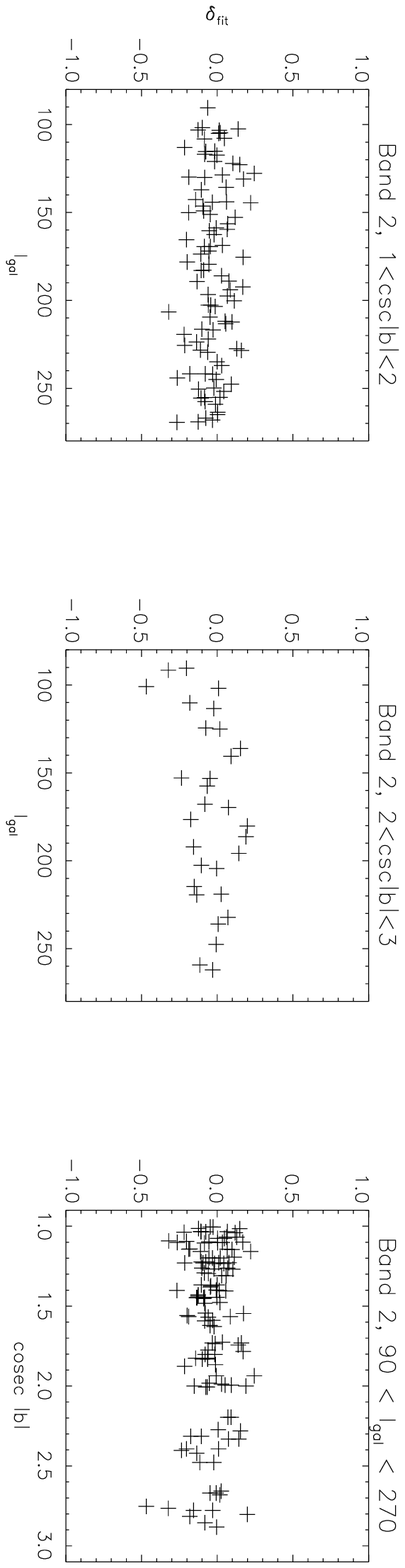}
\caption[]{ }
\label{f14}
\end{figure}

Fig.\ \ref{f14}: Scatter diagram for fit residuals $\delta_{\rm fit}\equiv
[\sqrt{C(0)}-\sqrt{C_{\rm fit}(0)}]/\sqrt{C(0)}$ for Eq.\ (\ref{e27}) for
Band 2 versus $l$ and $b$.

\begin{figure}
\centering
\leavevmode
\epsfxsize=1.0
\columnwidth
\epsfbox{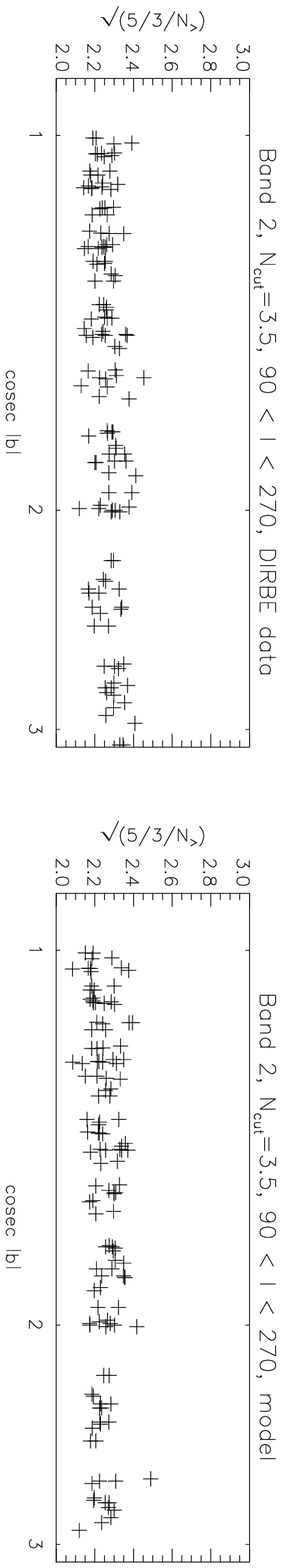}
\caption[]{ }
\label{f15}
\end{figure}

Fig.\ \ref{f15}: Effective clipping \Ncut vs.\ cosec$|b|$, according to
Eq.\ (\ref{e25})
for $p=2$ or
$B=0.3$. Left is DIRBE data and  right is simulated Galaxy model.

\begin{figure}
\centering
\leavevmode
\epsfxsize=1.0
\columnwidth
\epsfbox{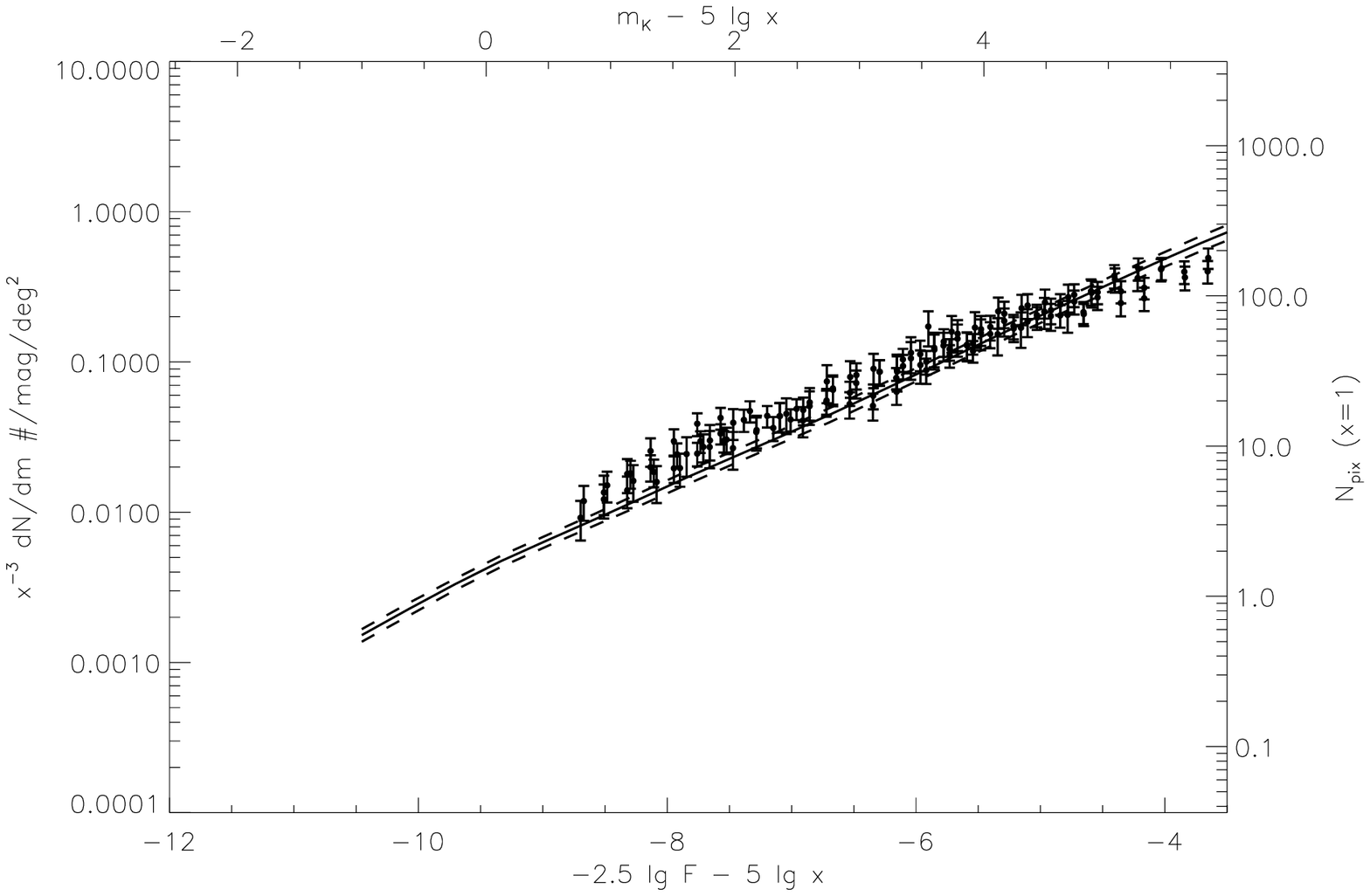}
\caption[]{ }
\label{f16}
\end{figure}

Fig.\ \ref{f16}: K-band DIRBE star counts in coordinates where a plane-parallel
Galaxy would be a
single line; $x=\cosec|b|$. Data are for $64 \times 64$ pixel patches with
$|b| \geq 20
\deg$ and
$90\deg < l < 270\deg$, with  Poisson errors shown for $N_{\rm pix} \geq
25$.  Confusion
noise  affects counts at  $m_K >5.5$. Lines show the model Galaxy:  solid
is mean
$x^{-3}dN/dm$ and dashes are the $\pm$1-sigma spread.

\begin{figure}
\centering
\leavevmode
\epsfxsize=1.0
\columnwidth
\epsfbox{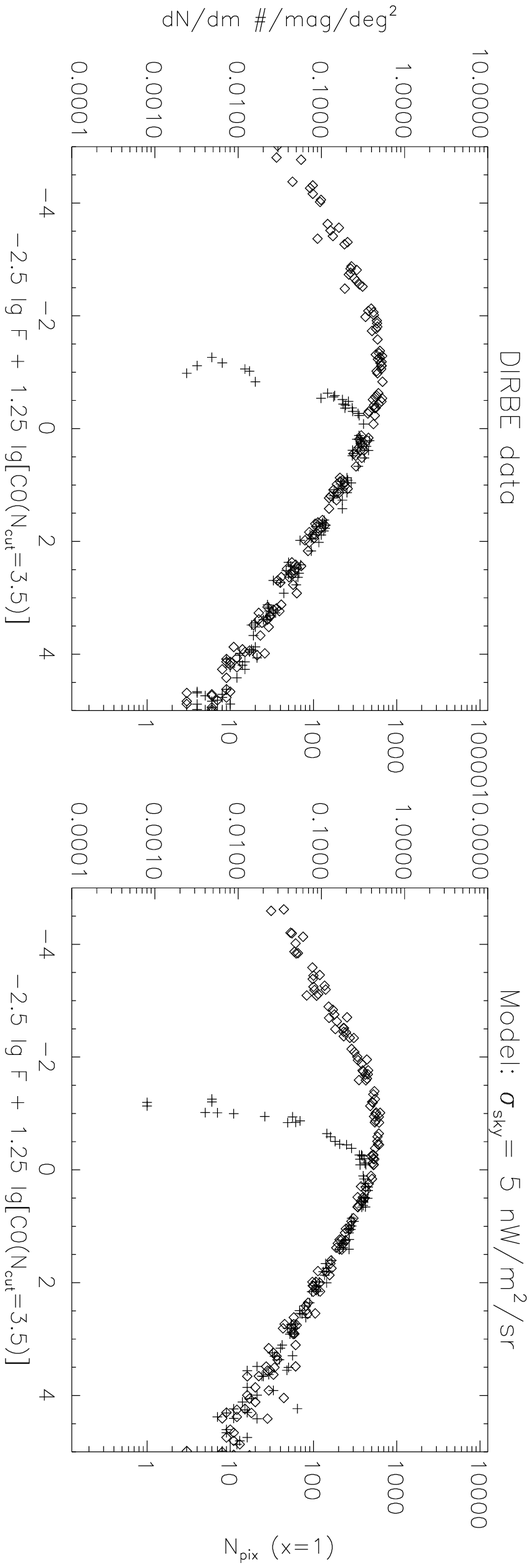}
\caption[]{ }
\label{f17}
\end{figure}

Fig.\ \ref{f17}: K-band star counts in the $64 \times 64$ patches outside
the Galactic
disk and away from the center, as a function of the absolute value of the
flux deviation
from the mean in units of $\sqrt{C(0)}$, for \Ncut = 3.5.  Left shows DIRBE
data and right
shows the simulated Galaxy and a CIB fluctuation of $\sigma_{\rm sky} = 5$
\nwm2sr
from Table \ref{t1}. Plus signs are for negative $F$ and diamonds for positive
$F$; they overlap
for small $|F|$.

\begin{figure}
\centering
\leavevmode
\epsfxsize=1.0
\columnwidth
\epsfbox{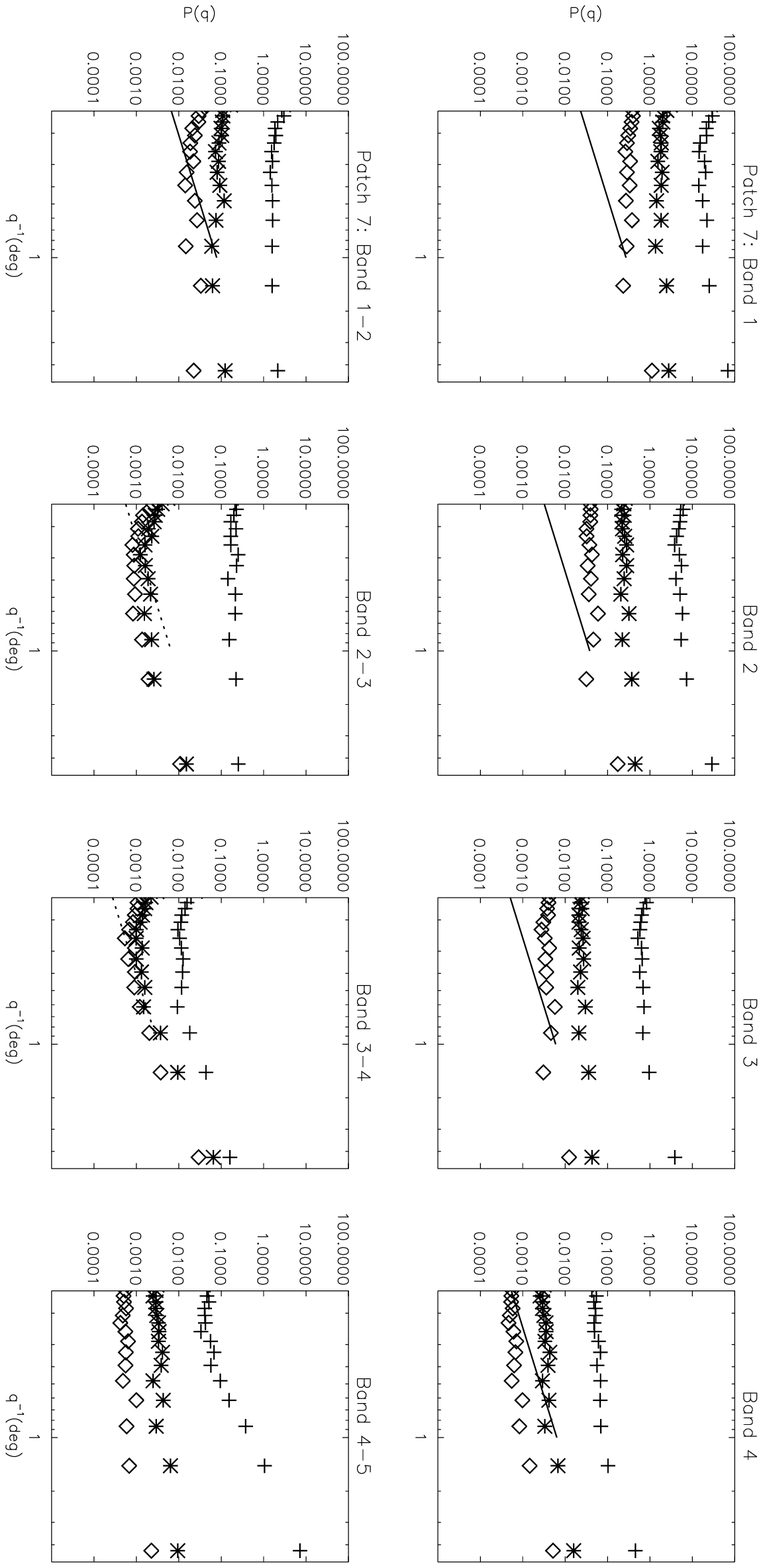}
\caption[]{ }
\label{f18}
\end{figure}

Fig.\ \ref{f18}: Near IR power spectra for Patch 7, well above the Galactic
and Ecliptic
planes.  Plus signs are before point source removal, asterisks show $N_{\rm
cut}=7$, and
diamonds show $N_{\rm cut}=5$. Solid lines show $P(q)$ expected if the CIB
power spectrum
scales as $q^{-1.3}$ for $a$ in Table \ref{t1}.

\begin{figure}
\centering
\leavevmode
\epsfxsize=1.0
\columnwidth
\epsfbox{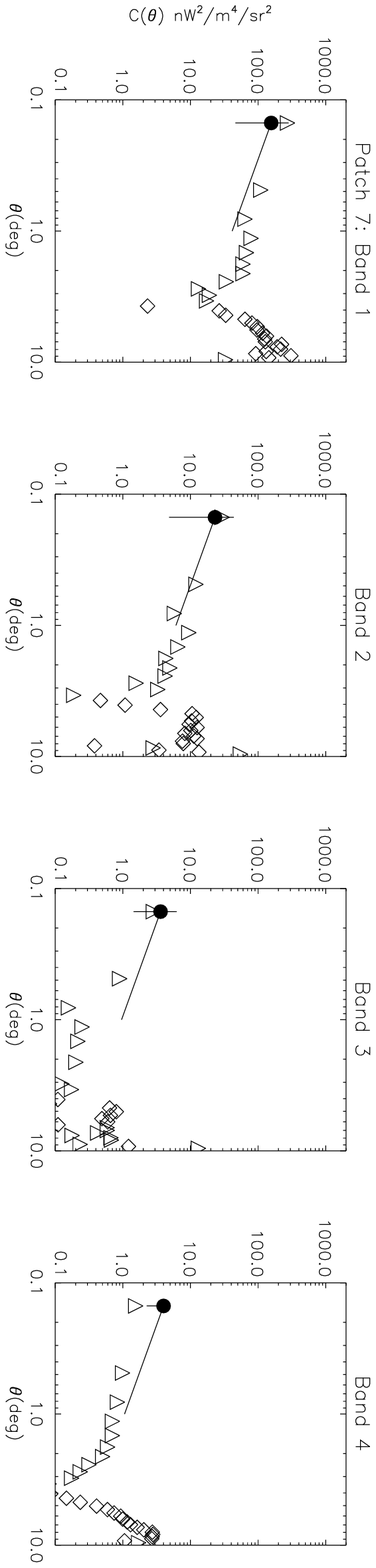}
\caption[]{ }
\label{f19}
\end{figure}

Fig.\ \ref{f19}: Angular correlation function evaluated for Patch 7 for
\Ncut = 3.5. The zero-lag value is plotted at $\theta=0.15^\deg$,
and numbers
below 0.1
\nw2m4sr are not shown. The filled circle shows $a$ from Table \ref{t1},
normalizing a CIB
correlation function with  $C(\theta) \propto \theta^{-0.7}$. Diamonds show the 
absolute value of the correlation function in the negative range.

\begin{figure}
\centering
\leavevmode
\epsfxsize=1.0
\columnwidth
\epsfbox{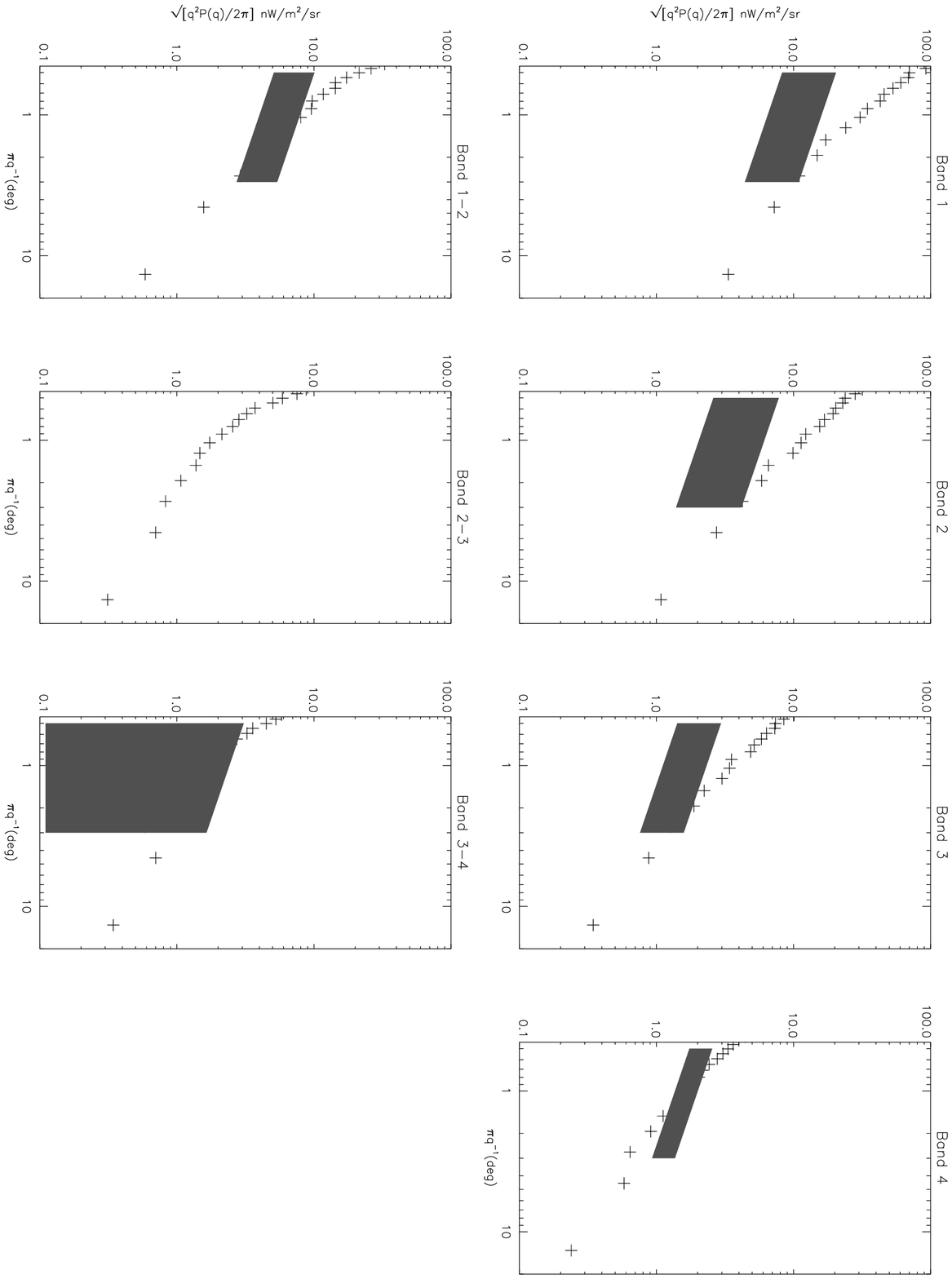}
\caption[]{ }
\label{f20}
\end{figure}

Fig.\ \ref{f20}: All-sky minimum fluctuation $\sqrt{q^2P(q)/2\pi}$ versus
$\pi/q$. Shaded
areas show the range expected from Table \ref{t1}, assuming the CIB power
spectrum
scales as
$q^{-1.3}$.

\begin{figure}
\centering
\leavevmode
\epsfxsize=1.0
\columnwidth
\epsfbox{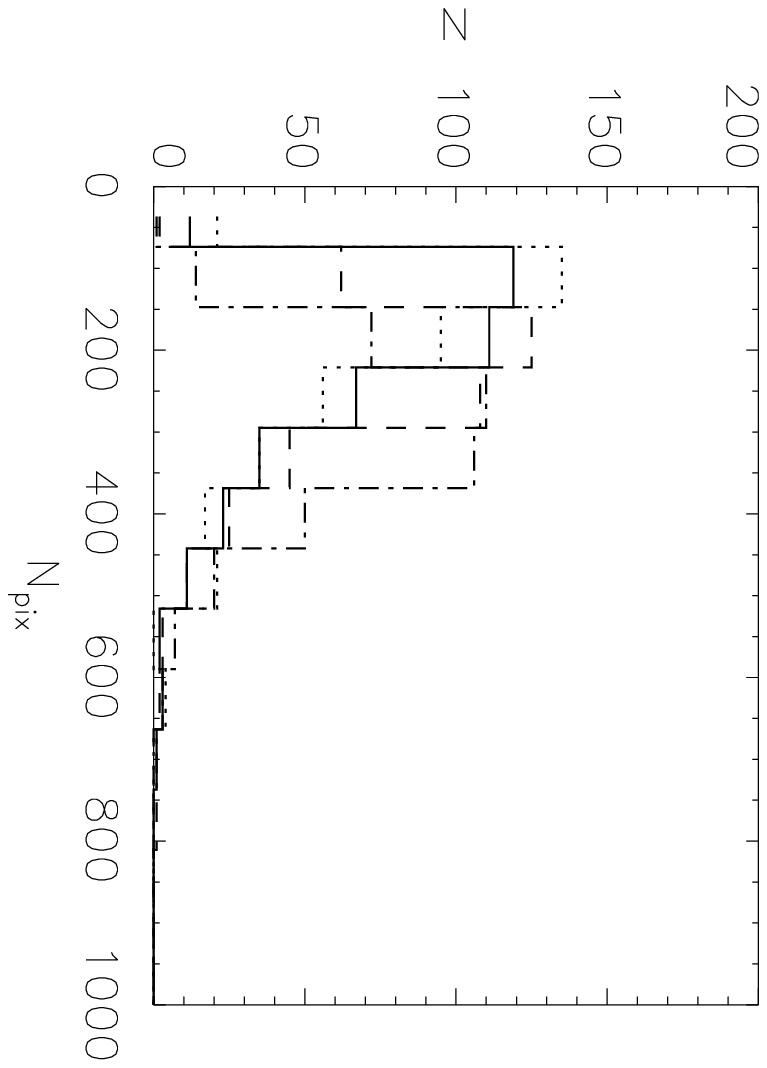}
\caption[]{ }
\label{f21}
\end{figure}

Fig.\ \ref{f21}: Histogram of pixels surviving clipping at \Ncut = 2 in the
mid- to far-IR
DIRBE bands  for 384 patches of $32\times32$ pixels. Solid line is Band 5,
dots Band 6,
dashes  Band 7, and dash-dots Band 8.

\begin{figure}
\centering
\leavevmode
\epsfxsize=1.0
\columnwidth
\epsfbox{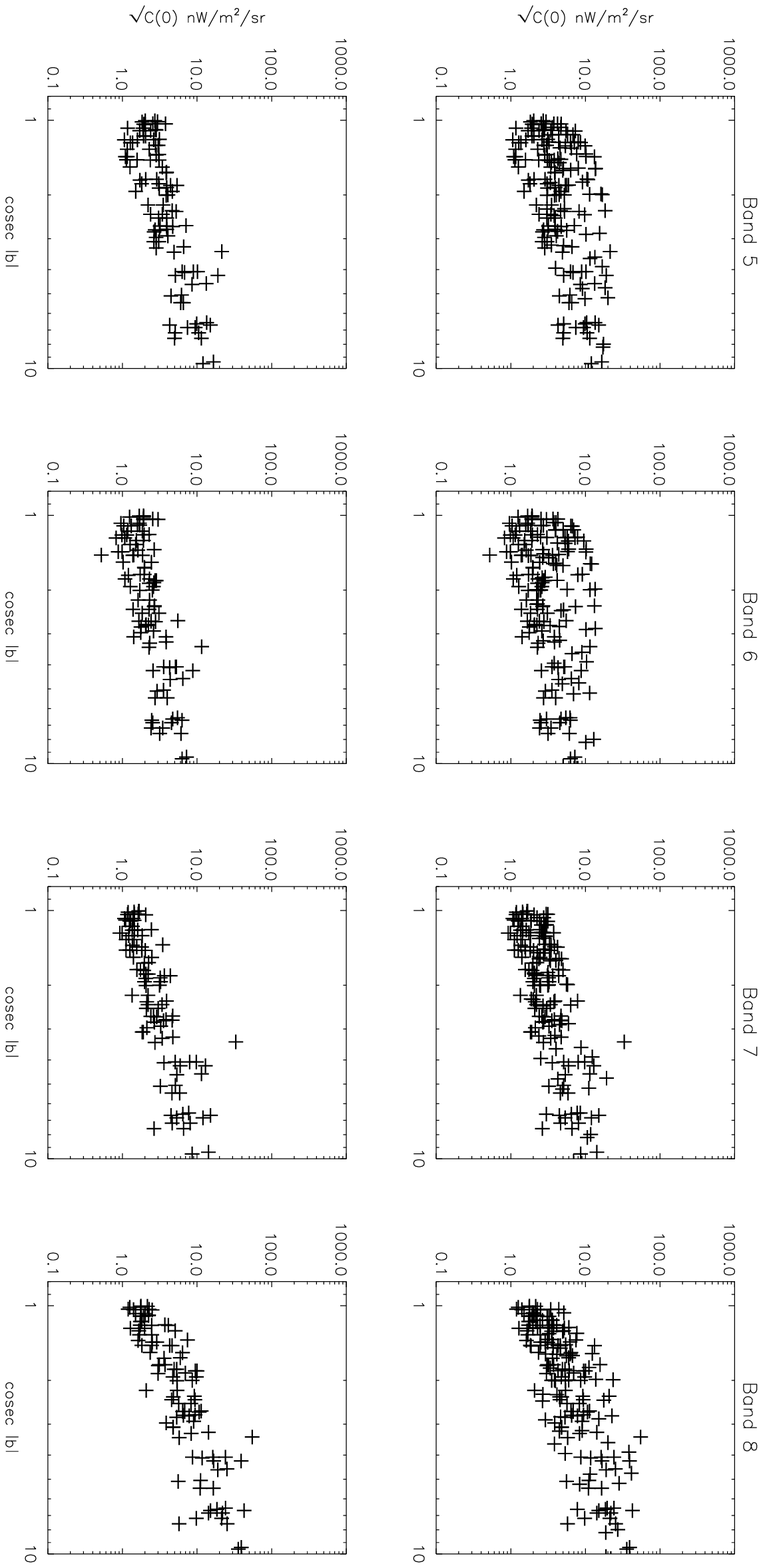}
\caption[]{ }
\label{f22}
\end{figure}

Fig.\ \ref{f22}: $\sqrt{C(0)}$ vs.\ cosec$|b|$ for Bands  5-8 and \Ncut =
2. Top panels are
90\deg\  $< l <$ 270\deg. Lower panels are $90\deg  < l < 270\deg$ and
$|\beta_{\rm Ecl}| > 25\deg$.

\begin{figure}
\centering
\leavevmode
\epsfxsize=1.0
\columnwidth
\epsfbox{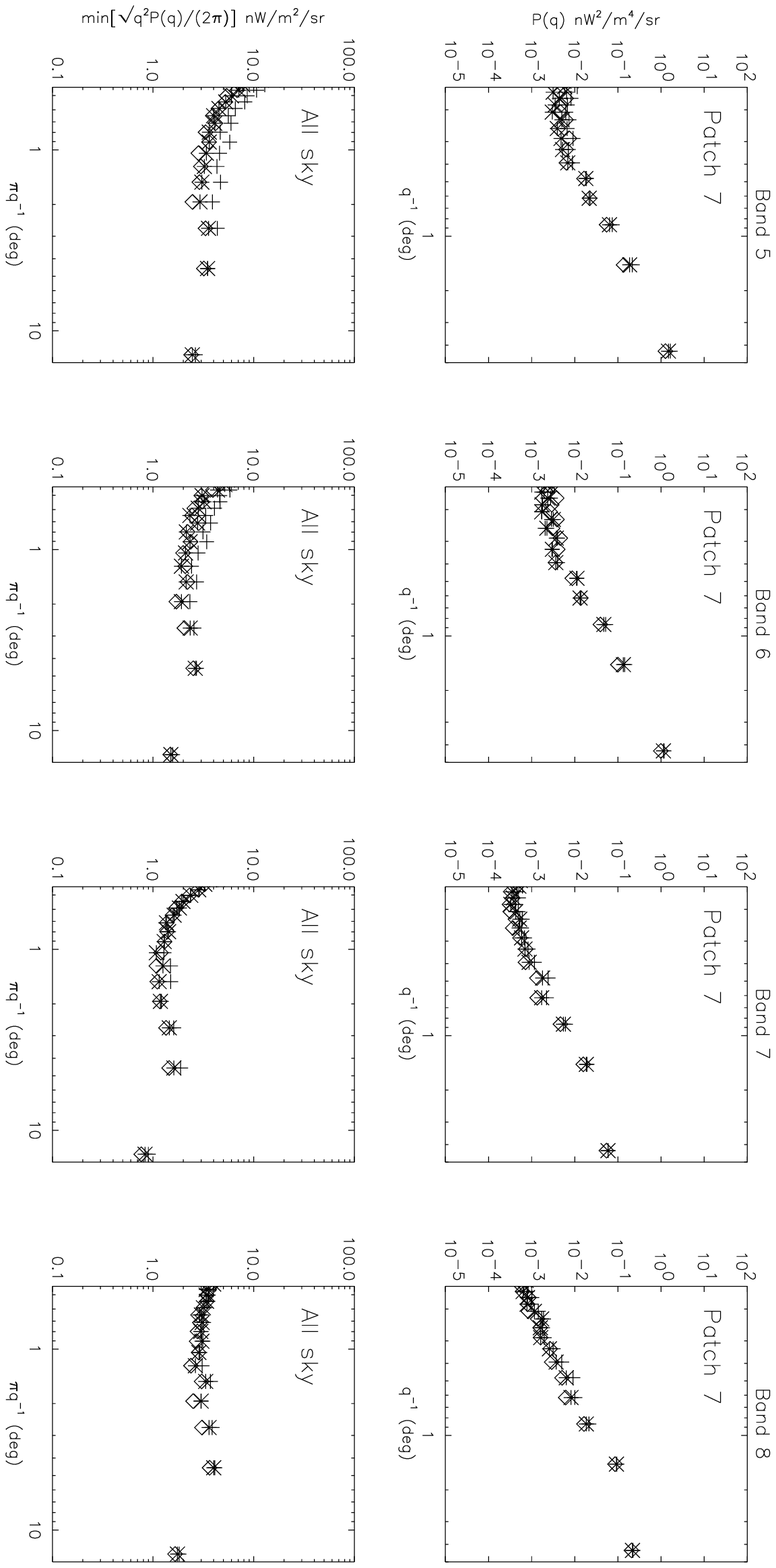}
\caption[]{ }
\label{f23}
\end{figure}

Fig.\ \ref{f23}: Upper panels are near-IR power spectra in for Patch 7.
Lower panels are
all-sky minimum fluctuations $\sqrt{q^2P(q)/2\pi}$ plotted vs.\ the scale
$\pi/q$.  Plus
signs are without clipping,  asterisks show $N_{\rm cut}=7$,  and diamonds
show $N_{\rm
cut}=5$.

Fig.\ \ref{f24}: Summary. Diamonds are values for $a$ with 92\%
uncertainties from Table
1. Triangles are for the color subtracted maps, shown at the mean
wavelength for the two
bands.  The $1-2$ limits are the left triangle and $3-4$  the right. Dashes
with
arrows are upper limits in Bands 4-8.

\newpage

\clearpage

\clearpage

\clearpage

\clearpage

\clearpage

\clearpage

\clearpage

\clearpage

\clearpage

\clearpage

\clearpage

\clearpage

\clearpage

\clearpage

\clearpage

\clearpage

\clearpage

\clearpage

\clearpage

\clearpage

\clearpage

\clearpage

\clearpage

\clearpage

\end{document}